# Accelerated Expansion as Predicted by an Ether Theory of Gravitation (*)

**Mayeul Arminjon**

**Abstract.** Cosmology is investigated within a new, scalar theory of gravitation, which is a preferred-frame bimetric theory with flat background metric. Before coming to cosmology, the motivation for an « ether theory » is exposed at length; the investigated concept of ether is presented in some detail: it is an elastically compressible fluid, and gravity is seen as Archimedes' thrust due to the macroscopic pressure gradient in that fluid; the construction of the theory, based on this concept, is explained; and the current status of the experimental confrontation is analysed, also in some detail. An analytical cosmological solution is obtained for a general form of the energy-momentum tensor. According to that theory, expansion is necessarily accelerated, both by the *vacuum* and even by matter. In one case, the theory predicts expansion, the density increasing without limit as time goes back to infinity. High density is thus obtained in the past, without a big-bang singularity. In the other case, the Universe follows a sequence of (non-identical) contraction-expansion cycles, each with finite maximum energy density; the current expansion phase will end by infinite dilution in some six billions of years. The density ratio of the present cycle (ratio of the maximum to current densities) is not determined by current cosmological parameters, unless a special assumption is made (« no cosmological time-dilation »). Since cosmological redshifts approaching $z = 4$ are observed, the density ratio should be at least 100. From this and the estimate of the current Hubble parameter, the time spent since the maximum density is constrained to be larger than several hundreds of billions of years. Yet if a high density ratio, compatible with the standard explanation for the light elements and the 2.7 K radiation, is assumed, then the age of the Universe is much larger still.

**Résumé.** On étudie la cosmologie correspondant à une nouvelle théorie de la gravitation; il s'agit d'une théorie scalaire bimétrique à référentiel privilégié, avec une métrique de référence plate. Avant d'en venir à la cosmologie, on expose de façon assez précise les raisons pour réintroduire un « éther »; on présente en quelque détail le concept d'éther envisagé – un fluide compressible dans lequel la gravitation est la poussée d'Archimède due au gradient macroscopique de pression – et la construction de la théorie à partir de ce concept; et l'on discute, assez en détail également, l'état actuel de la confrontation expérimentale. On obtient une solution cosmologique analytique pour une forme générale du tenseur énergie-impulsion. Selon cette théorie, l'expansion est nécessairement accélérée : elle l'est par le vide et même par la matière. Dans un cas, la théorie prévoit une expansion, avec un accroissement illimité de la densité lorsqu'on remonte vers le passé. On obtient ainsi de hautes densités dans le passé, tout en évitant une singularité du type « big-bang ». Dans l'autre cas, l'Univers suit une succession de cycles expansion-contraction distincts; pour chaque cycle, la densité maximum d'énergie reste finie. La phase actuelle d'expansion se terminera par une dilution infinie dans quelques six milliards d'années. Le rapport de densité du cycle actuel (rapport de la densité maximale à la densité actuelle) n'est pas déterminé par les valeurs actuelles des paramètres cosmologiques, sauf sous une hypothèse spéciale (« pas de dilatation-du-temps cosmologique »). Comme on observe des rougissements cosmologiques proches de $z = 4$, il faut que le rapport de densité soit au moins 100. Ceci, joint à l'estimation de la constante de Hubble, entraîne qu'il s'est écoulé plusieurs centaines de milliards d'années depuis la densité maximum. Si l'on admet une valeur élevée du rapport de densité, compatible avec l'explication usuelle pour les éléments légers et le fond continu à 2.7 K, alors l'âge de l'Univers est encore beaucoup plus grand.

---

## 1. INTRODUCTION

Due to its speculative character, cosmology is subject to radical changes. Thus, some years ago, it was thought that the cosmic expansion is slowed down. This may be seen from the very definition of a « deceleration parameter » $q$, for which a positive sign seemed *a priori* more plausible − and this sign was rather consistent with the then-available data.[1, 2] The main question was whether that deceleration would lead to a recontraction, or instead would remain compatible with a continued expansion for ever.[1-3] Within those times (in 1992), the author investigated, in an ether-based theory of gravitation,[4-6] some elementary consequences of the cosmological principle, and found that this theory says that $q \leq -4$. The then-accepted deceleration was one of the arguments against the steady-state theory,[7-8] which predicts $q = -1$, *i.e.* a constant expansion *rate*, as an immediate consequence of the perfect cosmological principle [Ref. [1], p. 461; Ref. [2], p. 268]. This did not encourage trying to publish a cosmology predicting that $q \leq -4$. Today the situation is different: the study of the relation redshift − luminosity for high-redshift supernovae suggests strongly that expansion is accelerated,[9-10] consistently (at least as regards the *sign* of $q$) with the ether theory and also with the steady-state theory. (The assumption of a homogeneous universe is so strong that, probably, one should not expect from a homogeneous cosmological model that it predicts the exact value of $q$ − the latter is, moreover, very difficult to measure in a model-independent way.) General relativity (GR) is able to *accommodate* such acceleration, by adjusting the value of the cosmological constant $\Lambda$, but GR cannot be said to *predict* an accelerated expansion. As is well-known, the physical meaning of $\Lambda$, though not entirely clear, has to do with the *vacuum* since, in the absence of any matter, a positive value of $\Lambda$ implies an accelerated expansion − thus giving physical properties to empty space, which should be the common feature of all « ethers ».

For these two reasons (the fact that it *predicts* an accelerated expansion, plus the fact that it is explicitly an ether theory), it seems interesting to see how cosmology would look in the ether theory of gravitation proposed in Refs. [4-6], and this is the aim of this paper. In *Section 2,* the motivation for building an « ether theory of gravitation » is explained, and the construction of this ether theory is presented in some detail. *Section 3* summarizes the current status of the confrontation of that theory with observation, starting from a general approximation scheme of the theory and applying this approximation scheme to the gravitational effects on light rays and to celestial mechanics. *Section 4* is devoted to the construction of cosmological models in the theory considered : first, the way in which expansion occurs in the most general metric of this theory is discussed; then the simplified equations that occur when one assumes (merely) that the Universe is *homogeneous* are derived. These equations are solved analytically in *Section 5:* it is found that the theory leads to two different possible scenarios for the evolution of a homogeneous universe − either a set of contraction-expansion cycles with non-singular bounce, or expansion with the density increasing without limit in the past. The analytical solution is used to evaluate some time scales in *Section 6.* In *Section 7,* the present results are discussed.

## 2. A PRESENTATION OF THE SCALAR ETHER THEORY OF GRAVITATION

The presentation chosen emphasizes the motivation and the way in which the equations are got. However, Hertz once said about Maxwell's theory (which Maxwell found with recourse to explicit « ether models »): « a theory is defined by the set of its equations. » Thus, one may take axiomatically the equations and definitions (2.5)-(2.15) and (2.18)-(2.19), which are the only ones that shall be used in the other Sections.

## 2.1 Why an ether theory ?

Following the operational philosophy of Einstein's famous 1905 paper on relativity, the concept of ether has been regarded with a growing suspicion during the first halfth of the past century, and is still regarded so by many physicists. However, other distinguished scientists who played an important role in the discovery of special relativity, namely Michelson, Larmor, Lorentz, and Poincaré, considered an ether until the end of their lives, although the two latter also adhered to relativity theory (see *e.g.* Poincaré[11] and Lorentz[12]). Since both of them were very great physicists, one may infer from this historical fact that the Lorentz-Poincaré ether theory – *i.e.,* the theory according to which the ether is an inertial frame in which Maxwell's equations are valid and such that any material object undergoes a « real » Lorentz contraction if it moves through the ether – must be compatible with special relativity. The compatibility of the [Lorentz-Poincaré rigid luminiferous] ether with special relativity was indeed explicitly acknowledged by Einstein in 1920.[13] The full compatibility of the Lorentz-Poincaré ether theory with special relativity in its usual, textbook form has been shown by Builder[14] and independently by Jánossy,[15] among others. It has been proved with a luxury of details by Prokhovnik,[16] and more recently by Pierseaux.[17] As emphasized by Bell[18] and by Prokhovnik,[19] the contraction assumption stated independently by FitzGerald and by Lorentz was not an *ad hoc* assumption since both FitzGerald and Lorentz derived this assumption from Heaviside's retarded potential of a uniformly moving charge, which indeed exhibits a « Lorentz-contracted » electric field. As noted by Duffy,[20] the Lorentz-Poincaré ether theory may be criticized only for the very reason that makes it physically equivalent to standard special relativity, namely the indetectability of its ether: since the theory is really Lorentz-invariant, any inertial frame may be indifferently choosed as the ether.[18] In this context, it is worth quoting Einstein's 1905 sentence[21] that led his followers (*i.e.,* most physicists) to reject the ether: « the introduction of a "luminiferous ether" will appear superfluous, insofar as, according to the conception developed here, neither does one introduce a "space in absolute rest" endowed with particular properties, nor is a velocity vector assigned to a point in empty space, at which electromagnetic processes take place. »[1] It can be seen that, even in 1905, Einstein did not reject the ether as leading to wrong physical laws, but rather considered the introduction of this concept as not needed any more, insofar as relativistic physical laws do not allow (by definition) to determine its velocity vector.

But special relativity does not involve gravitation. Therefore, the experimental support of special relativity does not, as such, provide much constraint on the theory of gravitation: one may define a very large class of « relativistic » modifications of Newton's gravity, by the condition that any of them should *reduce to special relativity when the gravitational field cancels.* (In GR, following Synge,[22] this may be defined as the case where the Riemann tensor cancels, in which case the Minkowski space-time is indeed obtained.) If one defines in this way the « relativistic » character of a theory of gravitation, nothing forbids that a such theory may have an *a priori* privileged reference frame – thereby violating, not only the explicit form of the principle of relativity (this is already the case for GR, as noted by Fock[23]), but its very spirit. Now the *physical reasons* to search for a such « ether theory of gravitation » come: A) from the wish to concile quantum physics with the theory of gravitation, and B) from some difficulties in GR itself.

**A)** Quantum mechanics and quantum field theory were built in *flat space-time* and actually need a preferred time coordinate. In flat space-time, one singles out the inertial time, and the fact that any inertial frame provides a different inertial time may be accommodated, because the standard quantization method (using a preferred time) leads naturally to Lorentz-invariant equations. But, in the curved space-time of GR, there is no way to prefer whatever time coordinate to another one, and this leads to the well-known « problem of time » in quantum gravity.[24] As a simple example, if one tries to directly extend a quantum wave equation from flat to curved space-time, by writing it in a covariant way, one is faced with the problem that covariant derivatives do not commute ; hence, even the Klein-Gordon equation can not be extended to curved space-time unambiguously in that way. On the other hand, the « quantum correspondence »

$$E \to +i\hbar\frac{\partial}{\partial t}, \quad p_j \to -i\hbar\frac{\partial}{\partial x^j}, \qquad (2.1)$$

that allows to pass from a classical Hamiltonian to a quantum wave equation, also becomes ambiguous in a curved space-time. In addition to the ambiguity in the time coordinate, this is because the Hamiltonian then contains mixed terms involving both the (4-)momentum and the (space-time) position : by (2.1), such mixed terms may lead to different, non-equivalent operators, depending on the order in which they are written.[25] As shown in the latter work, the Klein-Gordon wave equation is unambiguously extended to the situation with gravitation, as soon as one accepts to consider a *preferred-frame theory of gravitation* like that considered here. (And a preferred-frame equation is indeed obtained.) This extension is not merely formally correct, as could be some generalization of (2.1): it is consistent with the spirit of Schrödinger's wave mechanics.[25] Thus, it seems that it is not curved space-time as such that poses a serious problem with quantum theory, but specifically the absence of any « preferred space-time foliation » (in the words of Butterfield and Isham[24]), *i.e.* simply the reject of any ether. Moreover, quantum theory shows that what we call « *vacuum* » is endowed with physical properties (*cf.* the Casimir effect, which seems to be experimentally confirmed[26]), thus making an ether likely.

**B)** The difficulties that appear to exist in GR itself are the following ones : (**i**) The singularity occuring during gravitational collapse. Can a star really implode to a point singularity, or is it an unphysical prediction of some particular theories ? In any case, according to the scalar ether theory, the gravitational collapse of a spherical object in free fall leads instead to a bounce: [27] the object first implodes, as in GR, but then the implosion is stopped and is followed by an explosion.² (**ii**) The need to adopt a gauge condition in order to obtain explicit calculations. In the mathematical literature, the Einstein equations are usually interpreted as determining, not a Lorentzian metric $\gamma$ on a given space-time manifold V, but instead an equivalence class of couples (V, $\gamma$) modulo the equivalence relation « (V, $\gamma$) ~ (V', $\gamma$') if there is a diffeomorphism [an invertible mapping which is everywhere infinitely differentiable, together with the inverse mapping] of V onto V' that transforms $\gamma$ to $\gamma$'. »[28] However, this definition is difficult to handle in practice : in the physical literature, one considers the space-time (manifold) as given and the Einstein equations are supplemented with a gauge condition or coordinate condition. To the author's knowledge, there is not a precise link between these two ways of handling GR. In the scalar theory considered in this paper, the space-time manifold is given (moreover, it is equipped with a flat « background metric » $\gamma^0$) and there is no need for a gauge condition. (**iii**) The need for galactical dark matter in order to explain the motion of stars in galaxies (we refer here to the problem of the rotation velocities).[29] After many efforts, it seems that the identified candidates for the galactic dark matter are not found in sufficient amounts, and that the quest for dark matter is bifurcating towards the search for exotic particles yet to be discovered. Thus one may think that perhaps it is the standard gravitation theory that fails at these large scales.[30] In the scalar ether theory, there are preferred-frame effects and it is quite plausible that such effects might have a greater influence at large scales, because the long orbital periods (of the order of $10^8$ years) allow these effects to accumulate.

**2.2 Ether as a physical medium and the preferred reference frame**

*Warning.* – In the proposed theory, a concept of the ether as a physical medium is used to build equations which are then to be judged on their predicting ability. Most physicists of the 19th century after Fresnel considered an « ether, » whose first attribute was that it was the carrier of the light waves – since Fresnel and also Young had accumulated convincing evidence that light indeed behaved as waves, and because it was (and remains) difficult to think of waves without a medium to carry them. However, that imagined medium filling empty space could not be there just for light, it had also to be consistent with mechanics and the rest of physics. (As a relevant example, the ether considered by Lorentz and Poincaré was both the light-carrier and an inertial reference frame – although, in a later paper, Poincaré seemed to consider that the ether was like a fluid. [11]) This led to attribute some surprising and conflicting properties to the ether – it is well-known. Note that this « physical vacuum, » as some would call it now, still was named ether. Following or modifying ideas found in Romani's book, [31] we tried to build a concept of physical vacuum that is free from contradictions. We shall name it « ether » for short, and also because there is some analogy between our tentative concept and the ether concepts which were considered by Kelvin and by Poincaré. (Our tentative concept is summarized below.) But we do not believe that any concept may have a perfect correspondence with the physical world, in other words we do not adhere to a pure realism. We just feel that *there may be some truth in the concept described below*, and hope that *this truth may be deep enough so that a good theory of gravitation can be derived from it* (but certainly not a final theory, and our opinion is that no such theory will ever exist). Once a self-consistent set of equations is derived, thus the equations and definitions (2.5)-(2.15) and (2.18)-(2.19) for the present theory, it is essentially this set of equations that has to be assessed from the predictions that it leads to, and from the comparison of those predictions with observations. This does not preclude trying to interpret the various predictions by coming back to the model concept (here a fluid ether), but such interpretations are at a *different level* from the true physical assessment of the theory, which should be done by comparing observations with predictions deduced from the equations.

According to that concept, the ether or rather the « micro-ether » would be a *space-filling* perfect fluid, *continuous at any scale,* so that neither temperature nor entropy can be defined for that fluid, thus a *barotropic fluid* for which a one-to-one relationship between pressure and density must exist, $\rho_e = F(p_e)$. [31] Material particles should be *organized flows in that fluid,* such as vortices which may be everlasting in a perfect fluid. [31] Thus there would be nothing but ether. This very interesting idea should lead to a new theory of matter, and this is clearly a formidable task. But we can dispense ourselves from this task for the theory of gravitation, because there we shall use only an immediate consequence of this idea. On the other hand, the *average motion* of that fluid defines a preferred reference body, which plays the role of the Lorentz-Poincaré rigid ether. This is a new feature[4-5, 32] as compared with Romani's concept. [31] It avoids postulating an ether (physical vacuum) *plus* an independent absolute space, which would seem too much: the postulated ether should account for the inertial frames. It also avoids the absurdity which would result if one would assume the same continuous medium to be a fluid *and* perfectly rigid. Averaging the velocity field over space and time provides the way out of this paradox. Since any motion must be referred to some space and time, one formally starts from *space-time* as a flat (or possibly constant-curvature) Lorentzian manifold V, endowed with its natural metric $\gamma^0$. In what follows, we shall assume that it is flat, for simplicity. Then, the Lorentz-Poincaré ether is one particular inertial frame for the flat metric $\gamma^0$; this means that there is a global chart (coordinate system)[3] $\chi : X \mapsto (x^\mu)$ from V onto the arithmetic space $\mathbb{R}^4$, such that $(\gamma^0_{\mu\nu}) = (\eta_{\mu\nu}) \equiv \mathrm{diag}(1, -1, -1, -1)$ in the chart $\chi$, and such that the three-dimensional preferred reference body (manifold), denoted by M, is the set of the world-lines « $x^i = \mathrm{Const} = a^i$ ($i = 1, 2, 3$), $x^0$ arbitrary. » Thus each point of M may be defined by a 3-vector $\mathbf{x} \equiv (x^i)$. The ratio $t \equiv x^0/c$ is called the « absolute time, » where $c$ is a constant (the velocity of light). Let $\mathbf{u}'_e \equiv d\mathbf{x}/dt$ be the absolute velocity of the micro-ether and $p'_e$, $\rho'_e$ be the pressure and density fields of the micro-ether. According to the present ether concept, gravity is interpreted as due to the *macroscopic* gradient of the field of ether pressure (see § 2.3 below). Thus we first introduce macroscopic fields $\mathbf{u}_e$, $p_e$, $\rho_e$, obtained by

*volume*-averaging the foregoing fields in macroscopic, but finite domains of the space M. The volume measure used for that average is the Euclidean volume measure $V^0$, connected with the Euclidean metric $\mathbf{g}^0$ defined on the space M by $g^0{}_{ij} = \delta_{ij}$ in the chart $\chi$. Then, the assumption that M follows the average motion of the fluid micro-ether over space and time may be expressed formally as follows: *The time-average of* $\mathbf{u}_e$, at point $\mathbf{x} \in$ M and at time $t \in \mathbb{R}$ and in some fixed-length interval of time

$$\langle \mathbf{u}_e \rangle_{t,\mathbf{x}} \equiv \frac{1}{2T} \int_{t-T}^{t+T} \mathbf{u}_e(s,\mathbf{x}) \, ds, \qquad (2.2)$$

*is zero at any* $(t,\mathbf{x})$. [4] The global reference frame defined by all observers whose absolute velocity $d\mathbf{x}/dt$ is always zero, is the preferred reference frame of the theory, and shall be denoted by E. (Recall that $\mathbf{x}$ is the position in the preferred reference body M.)

**2.3 Gravity as Archimedes' thrust due to the macroscopic gradient of the ether pressure**

In this Subsection, we provisionally assume that mechanics can be formulated directly in terms of the Euclidean metric $\mathbf{g}^0$ and the absolute time $t$. (In § 2.4, we shall be led to assume that physical standards of length and time are affected by the field of ether pressure and we shall draw the consequence of that.) Since the average motion of the micro-ether defines the preferred reference frame of mechanics, this average motion itself does not obey a local mechanics. However, we may assume that the *microscopic* motion of the micro-ether *with respect to its mean rest frame* E, does obey mechanics, as well as the motion of material particles. The micro-ether being assumed to be a perfect fluid, it exerts only surface forces due to its pressure. The resultant force over a given particle occupying the domain $\omega$ may be reexpressed, using the divergence theorem, as a volume force depending on the gradient of the pressure of the micro-ether (provided the gradient may be considered constant in the vicinity of the particle): [4]

$$\mathbf{F}_g = -V(\omega) \operatorname{grad} p_e.$$

Thus, any material, « elementary » particle, supposed to occupy a finite volume, is subjected to this Archimedes thrust, just like a ball immersed in water (in this reasoning, we « forget » the fact that quantum, not classical mechanics, is usually admitted to apply at this scale). In order that this thrust be a universal force depending only on the mass of the material particles, it is sufficient to assume that the mass density inside any particle (averaged over the volume of the particle) is the same for all particles, say $\rho_p$. Then the force $\mathbf{F}_g$ may be rewritten as $\mathbf{F}_g = m\mathbf{g}$, with the gravity acceleration $\mathbf{g} = -(\operatorname{grad} p_e)/\rho_p$. [4, 27] But since we assume that matter particles should be just organized local flows in the micro-ether, $\rho_p$ is equal to $\rho_e$, the local density of the micro-ether. Then the former assumption is automatically satisfied insofar as $\rho_e$ is sufficiently uniform at the scale of matter particles. We obtain thus:

$$\mathbf{g} = -\frac{\operatorname{grad} p_e}{\rho_e}. \qquad (2.3)$$

Note that eq. (2.3) implies that $p_e$ and $\rho_e$ *decrease* towards the attraction. Moreover, since the gravitational force varies only over macroscopic distances, $p_e$ and $\rho_e$ must be actually the *macroscopic* pressure and density in the fluid ether. (See note 4.) More precisely, $\operatorname{grad} p_e$ must be the average of $\operatorname{grad} p'_e$ over a macroscopic volume, with $p'_e$ the microscopic pressure of the micro-ether. [4] This ensures that $\operatorname{grad} p_e$ is indeed uniform at the microscopic scale, and it also leaves the possibility that the other interactions, which vary over much shorter length scales, could be explained as combined effects of the field $p'_e$ on the matter particles, the latter being seen as organized flows. We see that gravitation can be *defined* by eq. (2.3), even in the case that the microscopic density $\rho'_e \equiv F(p'_e)$ would not be sufficiently uniform. Thus, at the scale of elementary particles, the universality of the gravitation force would be a question of definition. Moreover, the sequel of the development of the theory (Sect. 2.4) leads to admit that the macroscopic compressibility of the ether, defined in terms of the macroscopic fields, $K \equiv d\rho_e/dp_e$, is $1/c^2$ with $c$ the velocity of light, thus an extremely low compressibility. This implies that the microscopic compressibility, $K' \equiv d\rho'_e/dp'_e$, is also extremely small, hence one may expect that even $\rho'_e$ is nearly uniform over microscopic length scales.

Thus, in this ether model, the gravitational force is Archimedes' thrust that results from the spatial variation of the pressure of the fluid micro-ether over macroscopic scales, it is hence reduced to contact actions. Let us try to see why and how $p_e$ may vary over such scales. If this model describes correctly gravitation, $p_e$ has to decrease towards local concentrations of matter, eq. (2.3). Since matter is assumed to be made of organized flows in the micro-ether, one might hope that this influence of matter on the field $p_e$ should be itself deducible from local actions, but it would need that a precise description of matter in this ether model be available. However, if matter particles are made of vortices (or of complexes of several vortices[31]), then we may expect that indeed the presence of such vortices tends to decrease the pressure of the fluid: this is known from hydrodynamics, and it is a trivial observation in meteorology. Instead of attempting to derive the exact equation governing the ether pressure from microscopic considerations, we shall use a more phenomenological approach. If a concentration of matter makes $p_e$ decrease and thus produces a gravitational attraction, it follows that any motion of massive bodies (even a motion of one isolated body) causes a disturbance in the field of macroscopic ether pressure $p_e$, *i.e.* in the gravitational field. Since the ether is assumed compressible, this disturbance in the field $p_e$ cannot propagate instantaneously and instead should propagate as a pressure wave, so that an unsteady situation is obtained. Now Newtonian gravity (NG) propagates with infinite velocity, because it is the current mass density $\rho$ that determines the Newtonian potential and the gravity acceleration. If the fluid were incompressible, then the disturbances in the pressure field would indeed propagate instantaneously. Hence NG should correspond to the limiting case where the ether would be (macroscopically) incompressible, $\rho_e = \rho_{e0}$ = Const. From eq. (2.3), it follows immediately that Poisson's equation (div $\mathbf{g} = -4\pi G\rho$) and the whole of NG are exactly recovered in the latter case, if and only if the field $p_e$ obeys the following equation:

$$\Delta p_e = 4\pi G \rho \rho_{e0}, \qquad (2.4)$$

where $G$ is Newton's gravitational constant. In reality, no fluid is exactly incompressible but, since NG is known to be a very accurate theory of gravity, our fluid must be nearly incompressible, and NG, *i.e.* eq. (2.4), should be recovered asymptotically in situations where the effect of the compressibility can be neglected. This should be strictly the case in static situations, because then no propagation does occur. Hence we assume that eq. (2.4) holds true for a compressible fluid (thus with the space-varying field $\rho_e$ in the place of the constant $\rho_{e0}$) in the static case. We assume, moreover: (i) that the micro-ether is conserved, and hence that the macroscopic fields $\mathbf{u}_e, p_e, \rho_e$ obey the usual continuity equation of continuum mechanics; (ii) that the macroscopic motion of the micro-ether, defined by those fields, is a small wave motion with respect to the preferred frame [the latter being defined by the time-average (2.2)], and (iii) that this macroscopic motion obeys Newton's second law. On the basis of these assumptions, and by adapting the line of reasoning used in classical acoustics, we derived the following equation for the general, non-static situation: [4]

$$\Delta p_e - \frac{1}{c_e^2}\frac{\partial^2 p_e}{\partial t^2} = 4\pi G \rho \rho_e, \qquad (2.4\ \text{bis})$$

where we recall that $p_e$ and $\rho_e$ are related together by a one-to-one relation, and where

$$c_e = c_e(\rho_e) \equiv (dp_e/d\rho_e)^{1/2} \qquad (2.4\ \text{ter})$$

is the velocity of the macroscopic pressure waves − *i.e.,* of the *gravitational waves* − in the barotropic ether. (But that new theory differs from NG, hence $G$ cannot have exactly the same significance and value as in NG.) *In summary:* according to the concept of this theory, the gravitation force is directly due to the spatial variation of the ether pressure (or equivalently of the ether density) on macroscopic scales, but this variation, in turn, is due to the presence of matter and is hence influenced by the motion of matter – moreover, that influence propagates with a finite velocity. This is much like in Maxwell's theory, in which the Lorentz force is directly due to the all-pervading electromagnetic field, *i.e.* to the space-time variation of the electromagnetic potential, but the latter, in turn, is due to the presence and motion of charged particles – and it also involves a finite velocity of propagation.

**2.4 Principle of equivalence between the metric effects of motion and gravitation**

In the Lorentz-Poincaré version of special relativity, the Lorentz contraction is interpreted as a real contraction of all material objects in motion with respect to the ether. In the same way, the time period of a clock moving in the ether is interpreted as really dilated (furthermore, this « time-dilation » may be seen as a *consequence* of the Lorentz contraction[16]). Thus, there are real absolute effects of motion on the behavior of clocks and meters, *i.e.* absolute metric effects of motion. Similarly, in the scalar ether theory of gravitation, there are absolute metric effects of gravitation and, moreover, those latter effects are derived from the former ones. Gravitation is seen here as a variation in the ether density $\rho_e$ [see eq. (2.3)], and a variation in the « apparent » ether density indeed occurs in a uniform motion, due to the Lorentz contraction: for an observer having a constant velocity **u** with respect to the ether (but who is able to use the « true » simultaneity defined with the time of the ether frame), a given volume of ether has a greater volume, because his measuring rod is contracted in the direction **u**. Thus the apparent ether density is lowered in precisely the contraction ratio, *i.e.* the inverse of the Lorentz factor. This leads us to state the following assumption: [5, 27]

(A) *In a gravitational field, material objects are contracted, only in the direction of the field* **g** $=$ $-(\text{grad } p_e)/\rho_e$, *in the ratio* $\beta = \rho_e/\rho_e^\infty \leq 1$, *where $\rho_e^\infty$ is the ether density at a point where no gravity is present (far enough from massive bodies), and the clock periods are dilated in the same ratio.*

(Recall that $\rho_e$ decreases in the direction **g**. Therefore, $\rho_e^\infty$ is formally defined as the upper bound:

$$\rho_e^\infty(t) \equiv \text{Sup}_{\mathbf{x} \in M}\ \rho_e(t, \mathbf{x}), \qquad (2.5)$$

which must be finite.) This assumption is made for objects and clocks bound to the preferred frame; in general, one has to combine the metric effects due to motion and gravitation. Due to the space-contraction of measuring rods in the direction **g**, the « physical » space metric **g** (that measured with physical instruments) in the preferred frame E becomes a Riemannian one. The contraction occurs with respect to the Euclidean metric $\mathbf{g}^0$, introduced in § 2.2. The dilation of the clock periods in a gravitational field implies that the *local time* $t_\mathbf{x}$, measured by a clock at point **x** bound to the frame E, flows more slowly than the absolute time $t$:

$$dt_\mathbf{x}/dt = \beta(t, \mathbf{x}), \quad \beta \equiv \rho_e(t, \mathbf{x})/\rho_e^\infty(t) \quad (\beta \leq 1). \qquad (2.6)$$

Equation (2.6)$_1$ implicitly assumes that the absolute time $t$ is « globally synchronized ». Thus, it is assumed that the physical space-time metric $\gamma$ satisfies

$$\gamma_{0i} = 0 \ (i = 1, 2, 3) \tag{2.7}$$

in any coordinates ($x^\mu$) adapted to the frame E *and* such that $x^0 = ct$ with $t$ the absolute time. (« Adapted coordinates » are such that any particle bound to the given reference frame has constant space coordinates. [33]) Equation (2.7) implies[34] that we have also

$$\gamma_{ij} = - g_{ij}. \tag{2.8}$$

In any such coordinates, eq. (2.6) is equivalent to say that

$$\gamma_{00} = \beta^2. \tag{2.9}$$

Moreover, the notion that material particles are just organized flows in the micro-ether leads to assume that their velocity cannot exceed the velocity of pressure waves in the compressible ether, $c_e$, since otherwise they should be destroyed by shock waves. [4, 31] But SR must apply « locally » in the proposed theory, *e.g.* in the sense that the Lorentz contraction and time-dilation do affect moving objects, and also in the sense that dynamics of a particle is governed by the extended form of Newton's second law, eq. (2.16) below, involving the velocity-dependent inertial mass $m(v)$. This sets the other limit $c$ to the velocity of mass particles, hence one must have $c_e = c$ (everywhere at any time), and by (2.4 ter) this implies

$$p_e = c^2 \rho_e. \tag{2.10}$$

Thus, the waves of small disturbances in the ether pressure $p_e$, *i.e.* the *gravitational waves*, propagate with the velocity of light.

As a consequence of assumption (A), the line of reasoning leading to equations (2.3) and (2.4 bis) should be modified. It is first seen that one must now use, not the Euclidean space metric **g**$^0$, but instead the « physical » space metric **g** that obtains using physical length standards affected by gravitation. The compelling reason is that the assumed space contraction turns out to imply[5] that, as evaluated with the Euclidean metric, the macroscopic ether density has the space-uniform value $\rho_e^\infty(t)$ defined by eq. (2.5) [see eq. (4.2 bis) below], whereas our theory assumes a compressible ether and hence a space-nonuniform density $\rho_e$. Since it would appear absurd (at least ugly) to use physical length standards and « absolute » time standards, and also because local, physical standards are to be preferred anyway, the absolute time $t$ must also be replaced by the local time $t_\mathbf{x}$. It may then be inferred, though not rigorously proved,[5] that the same equations (2.3) and (2.4 bis) should now be valid in terms of the physical space metric **g** and the local time $t_\mathbf{x}$, and in replacing the Newtonian mass density $\rho$ by the density of mass-energy. Thus, the gravity acceleration is now defined by:

$$\mathbf{g} = -\frac{\mathrm{grad}_\mathbf{g}\, p_e}{\rho_e}, \qquad (\mathrm{grad}_\mathbf{g}\, \phi)^i \equiv g^{ij} \frac{\partial \phi}{\partial x^j}, \quad (g^{ij}) \equiv \mathbf{g}^{-1}, \tag{2.11}$$

and the field equation for the field $p_e$ is taken to be:

$$\Delta_{\mathbf{g}} p_e - \frac{1}{c^2}\frac{\partial^2 p_e}{\partial t_{\mathbf{x}}^2} = 4\pi G \sigma \, \rho_e, \tag{2.12}$$

where

$$\Delta_{\mathbf{g}} \phi \equiv \mathrm{div}_{\mathbf{g}} \, \mathrm{grad}_{\mathbf{g}} \, \phi = \frac{1}{\sqrt{g}}\frac{\partial}{\partial x^i}\left(\sqrt{g}\, g^{ij}\frac{\partial \phi}{\partial x^j}\right), \quad g \equiv \det(g_{ij}), \tag{2.13}$$

$$\frac{\partial}{\partial t_{\mathbf{x}}} \equiv \frac{1}{\beta(t,\mathbf{x})}\frac{\partial}{\partial t}, \tag{2.14}$$

and where $\sigma$ is the *mass-energy density of matter and nongravitational fields* in the preferred frame, precisely defined by [32]

$$\sigma \equiv (T^{00})_{\mathrm{E}} \qquad (x^0 = ct) \tag{2.15}$$

where **T** is the mass tensor (*i.e.* the energy-momentum tensor of matter and nongravitational fields, expressed in mass units) and with $x^0 = ct$ as the time coordinate.

**2.5 Dynamical equations**

Motion is governed by an extension of Newton's second law: [32] for a test particle, it writes

$$\mathbf{F}_0 + m(v)\mathbf{g} = D\mathbf{P}/Dt_{\mathbf{x}}, \tag{2.16}$$

where $\mathbf{F}_0$ is the non-gravitational (*e.g.* electromagnetic) force and $v$ is the modulus of the velocity **v** of the test particle (relative to the considered arbitrary frame F), the velocity **v** being measured with the local time $t_{\mathbf{x}}$ synchronized along the given trajectory[34] [$t_{\mathbf{x}}$ is given by eq. (2.6) if the preferred frame of the present theory is considered, thus if F = E] and its modulus $v$ being defined with the spatial metric **h** in the frame F (thus **h** = **g** if the preferred frame E is considered):

$$v^i \equiv dx^i/dt_{\mathbf{x}}, \quad v \equiv [\mathbf{h}(\mathbf{v},\mathbf{v})]^{1/2} = (h_{ij}\, v^i\, v^j)^{1/2}. \tag{2.17}$$

Moreover, $m(v) \equiv m(0).\gamma_v \equiv m(0).(1 - v^2/c^2)^{-1/2}$ is the relativistic inertial mass, $\mathbf{P} \equiv m(v)\mathbf{v}$ is the momentum, and $D/Dt_{\mathbf{x}}$ is the derivative of a spatial vector appropriate to the case where the Riemannian spatial metric **h** varies with time. In particular, with this derivative, Leibniz' rule for the time derivative of a scalar product $\mathbf{v}.\mathbf{w} = \mathbf{h}(\mathbf{v},\mathbf{w})$ is satisfied. [32, 35] With the gravity acceleration **g** assumed (in the preferred frame E) in the theory [eq. (2.11)], the extended form (2.16) of Newton's second law implies Einstein's motion along geodesics of metric $\gamma$, but only for a static gravitational field; however, (2.16) may be defined in any reference frame, and also for metric theories like GR: for those, one adds to the **g** of eq. (2.11) a certain *velocity-dependent term*, which gives geodesic motion in the general case. [35]

Equations (2.11) and (2.16) define Newton's second law in the preferred frame, for any mass particle. It is thereby defined also for a dust, since dust is a continuum made of coherently moving, non-interacting particles, each of which conserves its rest mass. It then mathematically implies, independently of the assumed form for the space-time metric $\gamma$ provided it satisfies $\gamma_{0i} = 0$ in the preferred frame, the following dynamical equation for the dust, in terms of its mass tensor $T^{\mu\nu} = \rho^* U^\mu U^\nu$ (with $\rho^*$ the proper rest-mass density and $U^\mu = dx^\mu/ds$ the 4-velocity): [36]

$$T^{\mu\nu}{}_{;\nu} = b^\mu. \tag{2.18}$$

Here $b_\mu$ is defined by

$$b_0(\mathbf{T}) \equiv \frac{1}{2} g_{jk,0} T^{jk}, \quad b_i(\mathbf{T}) \equiv -\frac{1}{2} g_{ik,0} T^{0k}. \tag{2.19}$$

(Indices are raised and lowered with metric **γ**, unless mentioned otherwise. Semicolon means covariant differentiation using the Christoffel connection associated with metric **γ**.) Equation (2.18), with the definition (2.19), is assumed to hold for any material continuum: accounting for the mass-energy equivalence, this is the expression of the universality of gravitation. Equation (2.18) is valid in any coordinates ($x^\mu$) that are adapted to the frame E and such that $x^0 = \phi(t)$ with $t$ the absolute time.

## 3. CURRENT STATUS OF THE CONFRONTATION WITH CLASSICAL TESTS

### 3.1 Main tests to be imposed on an alternative theory of gravitation

(**i**) The most important test for a theory of gravitation, and a less easy one than it might be believed, is that it should give nearly the same predictions as Newton's theory, in the great number of observational situations where NG is indeed extremely accurate. As it appears from the presentation above, the Newtonian limit is built in the very formulation of the scalar ether theory, but one has to make this statement more precise. (**ii**) The effects of gravitation on light rays, as predicted by GR, have been verified « to post-Newtonian (PN) accuracy, » *i.e.* to the accuracy of the first correction of GR to NG. [37] (**iii**) The explanation, by GR, of Mercury's residual advance in perihelion, also depends on the first correction to NG. [37] (**iv**) A viable theory should still predict that gravitation propagates with nearly the velocity of light and that, for a non-stationary insular matter distribution, gravitational energy is radiated towards outside, because the analysis of radio pulses received from binary pulsars such as the 1913+16 pulsar has shown that it is very likely to be the case. Indeed a model based on GR and the famous « quadrupole formula » reproduces extremely accurately the observed timing of successive pulses, including the long-term decrease in the time intervals (interpreted as resulting from the fall of the companion towards the pulsar, due to the energy loss). [38] (**v**) Finally, there are experimental results relevant to the verification of Einstein's equivalence principle (EEP). [37] As it appears from the equation for continuum dynamics, eq. (2.18) with the definition (2.19), the scalar ether theory agrees fully with the principle of the universality of gravitation. Moreover, eq. (2.18) coincides with the equation based on Einstein's equivalence principle (EEP), $T^{\mu\nu}{}_{;\nu} = 0$, in the case that the gravitational field does not depend on time – which is the case investigated in the analysis of tests of EEP. [37] For these two reasons, it is likely that the new theory passes the existing tests more or less as does GR. We shall hence focus on points (i)-(iv), but one day it should be possible to experimentally decide between EEP and the form of the equivalence principle that applies to the new theory, namely an equivalence between the absolute metric effects of motion and gravitation.

### 3.2 Asymptotic post-Newtonian approximation (PNA) of the scalar ether theory. Applications

When searching for an equivalent, in the scalar theory, of the PNA of GR, it has been choosed to start from the usual method of asymptotic expansion for a system of partial differential equations, because it has been mathematically tested in many different domains. [39-40] That method consists in expanding *all unknown fields* (see *e.g.* Refs. [41, 42] for illustrative applications in cases with several independent fields) in terms of a small parameter $\varepsilon$, which, at least conceptually, one should be able to make tend towards zero. This means that one should, at least formally, envisage a *family* ($S_\varepsilon$) of gravitating systems. [43] It has been found how to associate a such family with the gravitational system of interest, S (the solar system, say):[44] Since, in the scalar ether theory, the natural boundary value problem is the full initial-value problem, one applies a similarity transformation to the initial conditions fulfilled by the fields corresponding to the given system S. That transformation ensures that (i) the curved metric $\gamma_\varepsilon$ (for the system $S_\varepsilon$) tends towards the flat metric $\gamma^0$ as $\varepsilon \to 0$, and (ii) the matter fields (pressure $p_\varepsilon$, density $\rho_\varepsilon$ and velocity $\mathbf{u}_\varepsilon$) are of the same order in $\varepsilon$ as in the weak-field limit of NG. The latter is defined by an *exact* similarity transformation,[44, 45] *i.e.* one that applies to the fields, not merely at the initial time $t = 0$, but at any time. Thus, $p_\varepsilon$ is like $\varepsilon^4$, $\rho_\varepsilon$ is like $\varepsilon^2$, and $\mathbf{u}_\varepsilon$ is like $\varepsilon$, as $\varepsilon$ tends towards zero. The exact definition[44] of $\varepsilon$ implies that it satisfies

$$\varepsilon^2 \approx (U_\varepsilon)_{\max}/c^2, \qquad (U_\varepsilon)_{\max} \equiv \mathrm{Sup}_{\mathbf{x} \in M}\, U_\varepsilon(\mathbf{x}, t = 0) \qquad (3.1)$$

where $U$ is the Newtonian potential. This is the classical gravitational field-strength parameter,[46] thus the value $\varepsilon_0$ of $\varepsilon$ for the physically given system S is indeed small if the latter is a « weakly gravitating system » (as are most systems currently known with some accuracy): for the solar system, $\varepsilon_0^2$ is approximately $10^{-5}$ (reached at the centre of the Sun; however, values of $U/c^2$ outside the Sun are fractions of $10^{-6}$). If one changes units for system $S_\varepsilon$, multiplying the starting time unit by $\varepsilon^{-1}$ and the mass unit by $\varepsilon^2$, then $\varepsilon$ becomes proportional to $1/c$ and all matter fields are of order zero with respect to $\varepsilon$.[43, 44] This justifies to state expansions with respect to $1/c$, formally stated in the literature. (But all fields are expanded in the usual asymptotic expansion method.[44-45, 47] In contrast, only the gravitational field is expanded in the standard PNA.[23, 37, 46, 48-49]) Because it is actually $1/c^2$ that enters all equations, one may assume expansions in even powers of $1/c$.

It is then found that the equations of the zero-order approximation are those of NG.[43, 44] The next order is $1/c^2$ or $\varepsilon^2$, and this may be checked by assuming a first-order expansion in powers of $1/c$ (thus including the $1/c$ terms, unlike the even expansions): one finds that the complete equations of the first-order expansion also reduce to those of NG.[43] This means that, in weakly gravitating systems, the corrections of this theory to NG are of the order $\varepsilon^2$, thus very small. Hence *test (i) is fulfilled in the same way and to the same accuracy as in GR.*

Now the gravitational effects on light rays are: the gravitational redshift, the deflection, and the time-delay. The redshift is deduced, in GR, from the assumption that the frequency measured at the place of emission (proper frequency) is independent of the gravitational field. This assumption is consistent with the present theory as well, because the laws of nongravitational physics (*e.g.* the equations of Maxwell[6] and Klein-Gordon[25]) are firstly expressed in terms of the *physical* space-time metric $\gamma$. Hence it is indeed the frequency measured in terms of the local, physical time, that will appear as an output of the analysis of some quantum-mechanical oscillator (*e.g.* the Hydrogen atom). As for GR, however, it would be desirable to deduce the validity of the former assumption from a complete formulation of quantum mechanics in a gravitational field – but this is still lacking, also for GR (*cf.* § 2.1). The other elements that lead to the deduction of the three effects in GR are: the expression

$$\gamma_{00} = 1 - 2U/c^2 + \mathrm{O}(1/c^4) \qquad (3.2)$$

for the metric coefficient (for the redshift), and (for the other two effects) the fact that Schwarzschild's (exterior) metric is obtained for a spherical static body. Both facts are also true in the proposed theory, but there one has to account for the motion of the gravitating system through the « ether. » However, it has been proved that, even when accounting for that motion, eq. (3.2) remains valid, and also Schwarzschild motion is obtained, up to $\mathrm{O}(1/c^3)$ terms which play no role in the gravitational effects on light rays to PN accuracy. [50] As to test (ii), therefore: *to PN accuracy, the same effects are predicted on light rays as the standard effects of GR.*

*As regards test (iii)* (Mercury's advance in perihelion), the situation is interesting, less simple, and presently in a waiting position, because here *preferred-frame effects do intervene at the (first) PN level.* (The difference with light rays is that the velocity of light rays is O(*c*), of course, and this implies that less terms have to be kept for those, at a given approximation level.) The author has found that, if the velocity of the solar system through the « ether » is something like 300 km/s, as one expects if the 2.7 K blackbody radiation is at rest in the ether, then the contribution of the preferred-frame effects on Mercury's perihelion is larger than the Schwarzschild effect. Hence, if it *were* correct to take the « Newtonian » (first-approximation) contribution as *given independently of the theory of gravitation,* then the scalar theory would face a serious difficulty. But that is not correct. The point is that the masses of the planets, among other « Newtonian » parameters, are not measured (one cannot weigh a planet!), instead they are adjusted to best fit the observations. That the observational comparison involves a such *fitting* does not mean, of course, that celestial mechanics is not a predictive theory (*cf.* Neptune's discovery after Le Verrier's calculations, among other striking examples): the set of the input data, used in the fitting, does not necessarily contain all data, and various different sets may be tried. In Newtonian celestial mechanics, the fitting was formerly done by hand, essentially step by step by using the two-bodies solution with the orbital parameters as input values; with modern computers, one may make a global non-linear least-squares fitting. [51] Generally speaking, one has, in a given theory of gravitation, some equations of motion for the mass centers, in which the first-approximation masses and other unknown parameters appear, and one has to fit the observed mass-centers motion by the equations. Since those are theory-dependent, it follows that the optimal first-approximation masses $M_a$ are theory-dependent. More precisely, one may show that the « Newtonian » masses $M_a^N$, which are obtained by a fitting based on the first approximation *alone*, should differ from the $M_a$'s by second-approximation corrections. [52] This does not prove, of course, that the scalar theory correctly explains all minutes discrepancies between Newtonian celestial mechanics and observations in the solar system (Mercury's perihelion is merely the most significant discrepancy, and one that turns out to be explained if one adds the standard PN corrections of GR[53]). In order to judge the efficiency of that theory in celestial mechanics, one will have to make that global fitting referred to above. The current situation is that the PN equations of motion for the mass centers have been obtained, and a tentative algorithm for doing the fitting has been proposed. [52] In addition to the optimal first-approximation parameters, the output of the fitting will be the absolute velocity **V** of the solar system. If |**V**| were found negligible, this would put the scalar theory in a difficult situation. On the other hand, if a very good agreement were found for a significant value of |**V**|, that would be a decisive argument for an ether. Thus, the ether of the present theory is a falsifiable concept in the sense of Popper.

Finally, as to test (iv) relative to binary pulsars: in the « wave zone » (far away from the gravitating system, in the case of weakly self-gravitating systems), one expects that the scalar field has locally the structure of a plane wave, for which the derivatives with respect to *ct* have the same order of magnitude as the spatial derivatives. In that case, the relevant approximation of the scalar theory becomes a « post-Minkowskian » approximation (PMA) with wave equations for the gravitational potentials; the wave velocity is *c*, the velocity of light. In particular, for the first PMA, the field equation becomes just Poisson's equation with the wave operator in the place of the Laplacian. Now the exact theory admits a true local conservation for the energy (this is due to its lower covariance as compared with GR). [32] In the first PMA, the gravitational energy flux that appears in the energy balance is easy to evaluate at infinity, using multipole expansions of the retarded potential that solves the wave equation. If the global mass center of the isolated gravitating system is at rest in the « ether, » then dipole terms are eliminated and the time derivative of the global energy becomes very similar to the « quadrupole formula » (in particular, it contains neither monopole nor dipole, involves a $1/c^5$ factor, and corresponds to an energy *loss*), thus making it likely that binary pulsars data may be nicely fitted. What should be done now is (*a*) to justify the PMA as a consistent asymptotic scheme in the wave zone, as it has been done[44] for the PNA in the near zone (see above); and (*b*) to investigate the effect of a possible uniform translation of the mass center through the « ether. »

## 4. PRINCIPLES OF THE COSMOLOGICAL APPLICATION

### 4.1. General form of the metric, allowing expansion

The gravitational contraction of length standards, assumed to affect the "physical" space metric **g** in the preferred frame, occurs with respect to the Euclidean space metric $\mathbf{g}^0$, as explained in § 2.4. That flat metric is invariable (thus, its components do not depend on the absolute time $t$, in any coordinates adapted to the frame E). Since the gravitational contraction in the direction **g** is due to the *spatial* variation of the field $\rho_e$ (or equivalently that of the field $p_e$), there is still the possibility that **g** and $\mathbf{g}^0$ are related together by this contraction only up to a time-dependent factor $\alpha(t) > 0$.[27] This factor gives rise to a "cosmological" expansion or contraction of distances between objects bound to the preferred frame (and thus whose distances evaluated with $\mathbf{g}^0$ are invariable), as measured with the physical metric **g**. In the generic case where $\mathbf{g}(t, \mathbf{x})$ [or equivalently $\nabla \rho_e (t, \mathbf{x})$] $\neq 0$, we may express the assumption of a gravitational/cosmological contraction most simply in an 'isopotential' local coordinate system, *i.e.,* a local spatial coordinate system ($x^i$) such that, at the given time $t$, $x^1$ = Const. (in space) is equivalent to $\rho_e$ = Const, and such that the flat metric $\mathbf{g}^0$ is diagonal: $(g^0_{ij})$ = diag ($a^0_1$, $a^0_2$, $a^0_3$). Setting $R(t) = [\alpha(t)]^{1/2}$, our assumption gives then the following form to **g**:

$$(g_{ij}) = R(t)^2 \operatorname{diag}\,[\,(1/\beta)^2\,a^0_1,\ a^0_2,\ a^0_3], \qquad \beta \equiv \rho_e/\rho_e^\infty. \tag{4.1}$$

Hence the volume elements

$$\mathrm{d}V^0 \equiv \sqrt{g^0}\,\mathrm{d}x^1\,\mathrm{d}x^2\,\mathrm{d}x^3, \qquad \mathrm{d}V \equiv \sqrt{g}\,\mathrm{d}x^1\,\mathrm{d}x^2\,\mathrm{d}x^3$$

are related thus: [5]

$$\frac{\mathrm{d}V}{\mathrm{d}V^0} = \frac{\sqrt{g}}{\sqrt{g^0}} = R^3\,\frac{\rho_e^\infty}{\rho_e}. \tag{4.2}$$

The density $\rho_e$ defines the "amount" $\mathrm{d}m_e$ of ether in the volume element $\mathrm{d}V$: $\rho_e = \mathrm{d}m_e/\mathrm{d}V$. The density with respect to $\mathrm{d}V^0$ is hence

$$\rho_e^0 = \mathrm{d}m_e/\mathrm{d}V^0 = R^3\,\rho_e^\infty, \tag{4.2 bis}$$

which is space-uniform. But since the invariable metric $\mathbf{g}^0$ makes the ether a rigid body, [6] *the conservation of ether implies that $\rho_e^0$ must be constant*. [The conservation of ether is crucial in the derivation of the field equation in its first form, eq. (2.4 bis). [4]] Therefore, we get the relation

$$R(t)^3\,\rho_e^\infty(t) = \operatorname{Const} \equiv \rho_e^0. \tag{4.3}$$

Thus, the $R(t)^2$ factor in eq. (4.1) is not an independent factor and, since $\rho_e^\infty(t)$ is deduced from the field $\rho_e$ by eq. (2.5), the evolution of this factor is determined by the field equation (2.12) and the dynamical equation (2.18) − as we shall verify in the particular case of homogeneous cosmological models. Hence, *the presence of the $R(t)^2$ factor is necessary in the general case of the theory*. It means that, in addition to the « gravitational » space contraction in the direction **g** (which is due to the *spatial* heterogeneity of the field $\rho_e$), there is also a « cosmological » contraction of the measuring rods (if $R(t) > 1$ *i.e.* $1/R < 1$), which is due to the *temporal* heterogeneity of the field $\rho_e$. Now, in the case of gravitation, we have postulated a *time-dilation* as well, involving the *same ratio* $\beta$ as the space-contraction, just as is also the case for the effects of uniform motio − the FitzGerald-Lorentz space-contraction and the Larmor-Lorentz-Poincaré-Einstein time-dilation. The gravitational time-dilation is essential in the investigated ether theory: the gravitational time-dilation and space-contraction constitute the « ether-compatible form of the equivalence principle. » [5, 6, 27] It seems hence natural to expect that there also should be a cosmological time-dilation, involving the same ratio $R(t)$ as the cosmological space-contraction. [Note that it is actually $1/R$ that plays the same role as $\beta$, and recall that, as for gravitation, what is called here « space-contraction » is a contraction of measuring rods, hence a dilation (or « expansion ») of measured distances!] The space-time metric $\gamma$ would thus be, in coordinates adapted to E :

$$ds^2 = R(t)^{-2} \beta^2 (dx^0)^2 - g_{ij} dx^i dx^j, \qquad (4.4)$$

[with $x^0 = ct$ where $t$ is the absolute time, and with the $\beta^{-2}$ and $R(t)^2$ factors entering the relation between **g** and the invariable Euclidean space metric **g**$^0$, see eq. (4.1)], rather than

$$ds^2 = \beta^2 (dx^0)^2 - g_{ij} dx^i dx^j. \qquad (4.5)$$

However, it is eq. (4.5) that was implicitly assumed in the first version of this work. Comparing eqs. (4.4) and (4.5), we are led to assume a more general form:

$$ds^2 = R(t)^{-2n} \beta^2 (dx^0)^2 - g_{ij} dx^i dx^j, \qquad (4.6)$$

so that eqs. (4.4) and (4.5) correspond to $n = 1$ and $n = 0$, respectively. As we shall see, the equations are fairly the same independently of the value of *n*. Note that a time-scale factor independent of $R$, say $S(t)$ in the place of $R(t)^{-2n}$ in (4.6), can not be assumed, for it would not be determinable in this scalar theory. A power-law dependence seems then natural. Of course, eq. (2.6) is modified if one assumes (4.6), it becomes

$$dt_\mathbf{x}/dt = \beta(t, \mathbf{x})/R(t)^n, \qquad \beta \equiv \rho_e(t, \mathbf{x}) / \rho_e^\infty (t). \qquad (4.7)$$

Equation (4.4), *i.e.* $n = 1$, may appear as the most natural assumption within the investigated theory. Yet we note that, contrary to the gravitational space-contraction, which is in the **g** direction, the cosmological contraction is of a « conformal » kind: it affects all directions equally. But, in the present theory of gravitation, there is a flat space-time metric $\gamma^0$, whose line-element is defined, in coordinates adapted to the ether frame and with $x^0 = ct,$ by

$$ds^2 = (dx^0)^2 - g^0_{ij} dx^i dx^j. \qquad (4.8)$$

Now the « purely gravitational » space-contraction and time-dilation (*i.e.,* the case where the $R(t)$ factor is set equal to 1 in all equations of this Section) have the remarkable property that, in coordinates that are Galilean for the flat metric $\gamma^0$, the effective metric $\gamma$ is unimodular also, since

$$\gamma \equiv \det(\gamma_{\mu\nu}) = -1 \qquad (4.9)$$

in such coordinates [this is immediate from (4.2) and (4.6)]. This property is conserved in the « cosmological » case [*i.e.,* with $R(t)$], not for $n = 1$, but for $n = 3$. Thus, $n = 3$ is also a good candidate for the « natural » assumption. Once we have three « natural » candidates ($n = 0$, 1 and 3), it is clearly preferable to keep the general form (4.6). *This new form of the metric is valid in the general case for the present theory of gravitation:* we have adapted the form of the metric so that it is now valid *also* for cosmological problems. Of course, in problems at smaller scales, the time scales are much smaller so that the scale factor shall play a negligible role.

**4.2 Application of the homogeneity assumption**

The usual cosmological principle stipulates that, at a very large (cosmological) scale, the Universe is homogeneous and isotropic. [7] It is well-known that these two properties select a privileged reference frame, comoving with the « cosmic fluid », *i.e.,* with the average motion of matter at that very large scale. We shall assume merely the *homogeneity in the ether frame*. Thus it is assumed (i) that the spatial geometry in the ether frame E is spatially homogeneous; and (ii) that the mass tensor **T** does not depend on the spatial position in the frame E. Now it is obvious that, by construction, the geometry defined by metric (4.1) *cannot* be homogeneous in the neighborhood of any point **x** in the case where we actually derived (4.1), that is, the case where $\rho_e$ does vary in space in the neighborhood of **x**: with a space contraction in the ratio

$$\beta = \rho_e(t, \mathbf{x})/\rho_e^\infty(t),$$

the geometry is homogeneous if and only if $\beta$ is everywhere equal to unity. Note that eq. (4.1) still applies to that case (and this in any coordinates), and so does eq. (4.2). Assumption (i) is hence equivalent to assuming that the field $\rho_e$ (or the field $p_e$) is spatially uniform, *i.e.* $\rho_e(t, \mathbf{x}) = \rho_e^\infty(t)$ for all **x**, hence *there is no gravitational attraction* [$\mathbf{g} = 0$ in eq. (2.11)]. In that case, the metric (4.1) is automatically isotropic since it is proportional to the Euclidean metric:

$$g_{ij} = R(t)^2\, g^0_{ij}, \tag{4.10}$$

with $g^0_{ij} = \delta_{ij}$ in Cartesian coordinates for the Euclidean metric $\mathbf{g}^0$. Thus, *according to the scalar ether theory, the assumption of homogeneity implies the isotropy as far as spatial geometry is concerned*. Since $\beta = 1$ and keeping in mind eq. (4.1), the space-time metric (4.6) becomes

$$ds^2 = R(t)^{-2n}\,(dx^0)^2 - R(t)^2\, g^0_{ij}\, dx^i\, dx^j\,. \tag{4.11}$$

Moreover, $\beta = 1$ in eq. (4.7) means that the local (physical) time $t_\mathbf{x}$ flows uniformly in the ether frame and so represents a « cosmic time » which shall be denoted by $\tau$. Expressed in terms of $\tau$, the space-time metric $\gamma$ is thus the Robertson-Walker metric corresponding to a flat space:

$$ds^2 = c^2\, d\tau^2 - R(\tau)^2\, g^0_{ij}\, dx^i\, dx^j. \tag{4.12}$$

Therefore, the explanation of the observed redshift of the spectra emitted by distant galaxies as due to the expansion (the increase in $R$) is just as in GR (see *e.g.* Lachièze-Rey [3]). However, assuming $n \neq 0$ (which is more consistent with the rest of the theory than $n = 0$), the cosmic time $\tau$ does not coincide with the « absolute time » $t$ any more. Thus the latter becomes less directly attainable.

The field equation (2.12) was written in terms of the local time $t_\mathbf{x}$, which is now the uniform cosmic time $\tau$. Using (2.10) and the spatial uniformity of the field $\rho_e$, eq. (2.12) becomes

$$\ddot{\rho}_e + 4\pi G\sigma\,\rho_e = 0, \tag{4.13}$$

with $\sigma = \sigma(\tau)$ also since, according to assumption (ii), the mass-energy distribution is homogeneous. In eq. (4.13) and henceforth, *the upper dot means differentiation with respect to the cosmic time $\tau$.* Note, however, that although the time differentiation is with respect to $\tau$ in (4.13), $\sigma$ is the $T^{00}$ component of the mass tensor with $x^0 = ct$ as the time coordinate [eq. (2.15)]. Substituting $\rho_e = \text{Const}/R^3$ [eq. (4.3)] into (4.13) and rearranging, we get:

$$\frac{\ddot{R}}{R} = \frac{4\pi G}{3}\sigma + 4\frac{\dot{R}^2}{R^2}, \tag{4.14}$$

whence, insofar as $\sigma \geq 0$,

$$q \equiv -\frac{R\ddot{R}}{\dot{R}^2} = -\left(\frac{4\pi G}{3}\sigma\frac{R^2}{\dot{R}^2} + 4\right) \leq -4. \tag{4.15}$$

Thus, expansion is *necessarily* accelerated, according to the ether theory. Furthermore, the higher the matter density $\sigma$, the *more* expansion is accelerated. This is in striking contrast with GR which, for a perfect fluid with proper rest-mass density $\rho$ and pressure $p$, leads to [Ref. [(2)], p. 191]

$$\frac{\ddot{R}}{R} = -\frac{4\pi G}{3}\left(\rho + \frac{3p}{c^2}\right) + \frac{\Lambda}{3}c^2, \tag{4.16}$$

instead of eq. (4.14); thus an increase in the energy density tends to decelerate the expansion, according to GR. It is often commented on that point as if it were a normal gravitational effect, *i.e.,* matter attracts matter and so slows down expansion. But we are here in an (assumed) homogeneous Universe and hence there is no gravitation in the Newtonian sense. (And which direction could take the gravitational attraction in an isotropic Universe?) [8] Note that, in the present theory, eqs. (4.14) and (4.15) are valid for any material medium, independently of the state equation. On the other hand, the contribution $\lambda = 4\dot{R}^2/R^2$ in the r.h.s. of eq. (4.14) is there independently of the presence or absence of matter, and may be interpreted as the « *vacuum* contribution, » just as the contribution $\Lambda c^2/3$ in eq. (4.16) of GR. The latter contribution could be made equal to the $\lambda$ term if one would assume $\Lambda = \Lambda(\tau) = 12\,H^2/c^2$ with $H(\tau) = \dot{R}/R$ the Hubble parameter.

Let us now rewrite the dynamical equation (2.18) for a *homogeneous* Universe. We take Cartesian space coordinates for the Euclidean metric $\mathbf{g}^0$ and compute the l.h.s. with the identity

$$T_\mu{}^\nu{}_{;\nu} = \frac{1}{\sqrt{-\gamma}}\left(\sqrt{-\gamma}\,T_\mu{}^\nu\right)_{,\nu} - \frac{1}{2}\gamma_{\lambda\nu,\mu}\,T^{\lambda\nu} \qquad (\gamma \equiv \det(\gamma_{\lambda\nu})), \tag{4.17}$$

using $x'^0 \equiv c\tau$ as the time coordinate [since eq. (2.18) admits the change $x'^0 = \phi(x^0)$] and hence the Robertson-Walker form (4.12) for the space-time metric $\gamma$. We first get $-\gamma = R(\tau)^6$. Moreover, the material homogeneity (ii) implies that $T_\mu{}^\nu{}_{,j} = 0$. The spatial part of eq. (2.18) writes then:

$$\frac{1}{R^3}\left(R^3 T_i{}^0\right)_{,0} = -RR_{,0}\,T^{0i}, \tag{4.18}$$

and since $T_i{}^0 = -R^2\,T^{0i}$, it is still equivalent to

$$T^{0i}{}_{,0}/T^{0i} = -4\,R_{,0}/R \quad \text{(no sum over } i\text{)}, \qquad \text{i.e. } T^{0i}(\tau) = C^i/R(\tau)^4 \tag{4.19}$$

with $C^i$ a constant. Note that the spatial vector with components $T^{0i}$ (and which represents the density of the energy flux) would have to be zero if the matter distribution were assumed *isotropic* in the preferred frame. A non-zero vector **C** means, at any given time, a spatially uniform energy flux vector. That is, matter may have a spatially uniform motion with respect to the preferred frame, the time evolution of this motion being determined, according to eq. (4.19), by that of the scale factor $R(\tau)$. This is *the momentum conservation for an expanding (or contracting) homogeneous Universe.* It shows in particular that matter was and will remain at rest in the preferred frame (*i.e.,* $T^{0i}(\tau) = 0$ for any $\tau$) if it is at rest at some given time. As to the time part of eq. (2.18), it gives in the same way:

$$\frac{1}{R^3}\left(R^3 T_0^{\ \nu}\right)_{,\nu} = 0. \tag{4.20}$$

Setting $\varepsilon \equiv T_0^{\ 0}$ [which is invariant by a time change $x'^0 = \phi(x^0)$], the material homogeneity (ii) allows to rewrite this as

$$R^3 \varepsilon = \text{Const.} \equiv \varepsilon^{(0)}, \tag{4.21}$$

whence

$$\dot{\varepsilon}/\varepsilon = -3\dot{R}/R = -3H(\tau). \tag{4.22}$$

Now $\varepsilon = T_0^{\ 0}$ is the density of material mass-energy with respect to the *physical* volume measure $dV$, *i.e.* $\varepsilon = de/dV$.[9] Equation (4.21) means that the density of material mass-energy with respect to the invariable Euclidean volume measure $dV^0$ is the constant $\varepsilon^{(0)}$. Indeed, eqs. (4.2) and (4.3) (with now $\rho_e^\infty = \rho_e$) show that

$$\frac{de}{dV^0} = \varepsilon \frac{dV}{dV^0} = R^3 \varepsilon \equiv \varepsilon^{(0)}. \tag{4.23}$$

Hence, equation (4.21) expresses the *conservation of the material mass-energy* in the context of homogeneous cosmological models. That true *conservation* is thus a derivable consequence of the proposed theory. Together with eq. (4.3), *i.e.* the *conservation of ether,* eq. (4.21) still shows that

$$\varepsilon = C\rho_e \tag{4.24}$$

with $C$ a constant. Coming back to the field equation (4.13), recall that $\sigma \equiv T^{00}$ with $x^0 = ct$ as the time coordinate. Denoting with a prime, here only, the components when the time coordinate is

$$x'^0 \equiv c\tau,$$

we have $T'^{00} = T'_0^{\ 0} \equiv \varepsilon$ since $\gamma'_{00} = 1$ by eq. (4.12), hence by (4.7) [where now $t_\mathbf{x} = \tau$ and $\beta = 1$]:

$$\sigma = (dt/d\tau)^2 T'^{00} = R^{2n} \varepsilon. \tag{4.25}$$

Using eqs. (4.24) and (4.25), we rewrite eq. (4.13) as

$$\ddot{\varepsilon} + (4\pi G \varepsilon R^{2n})\varepsilon = 0. \tag{4.26}$$

Eliminating $R$ by eq. (4.21), we finally get

$$\ddot{\varepsilon} + \alpha' \varepsilon^{2-2n/3} = 0, \quad \alpha' \equiv 4\pi G\left[\varepsilon^{(0)}\right]^{2n/3}. \tag{4.27}$$

This is a non-linear oscillator equation and represents the seed for an oscillatory behaviour of the energy density $\varepsilon$ and consequently also for the scale factor $R$ [eq. (4.23)]. That *oscillatory behaviour of the Universe* will be demonstrated analytically below.

## 5. ANALYTICAL COSMOLOGICAL SOLUTION

The differential equation (4.27) can be solved analytically. (The case $n = 0$ was already encountered in the study of gravitational collapse in the ether theory.[27]) It admits the first integral (for $n \neq 9/2$):

$$\dot{\varepsilon}^2 + \alpha \varepsilon^m = \text{Const.} \equiv A, \qquad \alpha \equiv 2\alpha'/m = 24\pi G \left[\varepsilon^{(0)}\right]^{2n/3} / (9 - 2n), \qquad (5.1)$$

where the exponent $m$ is related to the time-dilation exponent $n$ by

$$m = 3 - 2n/3.$$

We have thus the following relation between the signs:

$$\text{sgn}(\alpha) = \text{sgn}(m) \qquad \text{(for } n \neq 9/2\text{)}. \qquad (5.2)$$

Recall that $n = 0$ means no cosmological time-dilation, which was an implicit assumption in the first version of this work, and that two values, $n = 1$ and $n = 3$ (thus $m = 7/3$ and $m = 1$) appear "potentially privileged" in the theory considered, although other values of $n$ are quite possible also (§ 4.1). Since both $\varepsilon$ and $R$ are positive, eq. (4.22) shows that the sign of $\dot{\varepsilon}$ is opposite to that of $\dot{R}$, hence $\dot{\varepsilon} > 0$ means contraction and $\dot{\varepsilon} < 0$ means expansion. We have from (5.1):

$$\frac{d\tau}{d\varepsilon} = \frac{\pm 1}{\sqrt{A - \alpha \varepsilon^m}} \qquad (5.3)$$

when the square root is defined, whence the analytical solution. Let us first consider an interval of time with expansion, *i.e.* $\dot{\varepsilon} \leq 0$, or equivalently $d\tau/d\varepsilon \leq 0$, thus the minus sign in (5.3). The solution to (4.27) is then given, in inverse form, by integrating (5.3):

$$\tau = \tau_{\text{ref}} - F(\varepsilon), \qquad (5.4)$$

where $\tau_{\text{ref}}$ is the value of the cosmic time at which the value of $\varepsilon$ is $\varepsilon_{\text{ref}}$, and where

$$F(\varepsilon) \equiv \int_{\varepsilon_{\text{ref}}}^{\varepsilon} \frac{d\xi}{\sqrt{A - \alpha \xi^m}}. \qquad (5.5)$$

This is certainly defined if the square root exists and does not cancel in the interval [$\varepsilon_{\text{ref}}$, $\varepsilon$]. Now, if $A/\alpha$ is positive, the radicand cancels at one value of $\xi$, which we shall denote by $\varepsilon_{\text{max}}$. Indeed it will be proved later that, in that case, $\varepsilon_{\text{max}}$ is a true maximum of the solution $\varepsilon = \varepsilon(\tau)$. From (4.22), we get

$$\varepsilon_{\text{max}} = (A/\alpha)^{1/m} = \varepsilon_0 \left(1 + \frac{9 H_0^2}{\alpha \varepsilon_0^{m-2}}\right)^{1/m}. \qquad (5.6)$$

(If $A/\alpha$ is negative, then the radicand cannot cancel.) *Index* 0 *means henceforth quantities taken at cosmic time $\tau_0$ ; in the numerical applications, we shall take for $\tau_0$ the current epoch.* In the special case $m = 0$, *i.e.* $n = 9/2$, eq. (5.4) remains true although we must write

$$\dot{\varepsilon}^2 + \alpha \operatorname{Log} \varepsilon = \operatorname{Const.} \equiv A, \qquad \alpha \equiv 2\alpha' = 8\pi G \left[\varepsilon^{(0)}\right]^3 > 0 , \qquad (5.1)'$$

$$F(\varepsilon) \equiv \int_{\varepsilon_{\text{ref}}}^{\varepsilon} \frac{d\xi}{\sqrt{\alpha \operatorname{Log}(\varepsilon_{\max} / \xi)}}, \qquad (5.5)'$$

$$\varepsilon_{\max} = \exp(A/\alpha) = \varepsilon_0 \exp(9H_0^2 \varepsilon_0^2 / \alpha), \qquad (5.6)'$$

instead of (5.1), (5.5) and (5.6). Moreover, in contrast with the case $m \neq 0$, the value (5.6)' is defined for whatever sign of $A/\alpha$.

Now the *big difference* between $n = 0$ (*i.e.* $m = 3$) and any $n \neq 0$ is the presence of the "hidden parameter" $\varepsilon^{(0)}$ [eq. (4.23)] in coefficient $\alpha$ if $n \neq 0$. This means that, *if $n \neq 0$ and $A/\alpha > 0$*, the density ratio $\varepsilon_{\max}/\varepsilon_0$ cannot be calculated as function of the current values of the mean energy density and the Hubble parameter, $\varepsilon_0$ and $H_0$, whereas it can if $n = 0$. In rude words: *the density ratio $\varepsilon_{\max}/\varepsilon_0$ becomes a free parameter of the model if $n \neq 0$* – whereas, if $n = 0$ (as was assumed in the first version of this work), then

$$\frac{\varepsilon_{\max}}{\varepsilon_0} = \left(1 + \frac{27 H_0^2}{8\pi G \varepsilon_0}\right)^{1/3} \qquad (n = 0). \qquad (5.7)$$

**5.1 Case $n \leq 9/2$ : contraction-expansion cycles with non-singular bounce**

In the case $n \leq 9/2$, *i.e.* $m \geq 0$, we have $\alpha > 0$ [eqs. (5.2) or (5.1)'], and for $n < 9/2$ we have therefore $A > 0$ [eq. (5.1)]. It follows that the radicand does cancel for the value (5.6) [or (5.6)']. If $m > 0$, eq. (5.3) becomes, in units such that $A = \alpha = 1$ (so that $\varepsilon_{\max} = 1$):

$$d\tau/d\varepsilon = \pm [1 - (1-x)^m]^{-1/2} = \pm [mx + O(x^2)]^{-1/2}, \qquad x \equiv 1 - \varepsilon \to 0^+, \qquad (5.8)$$

so that the integral (5.5) is finite at $\varepsilon = \varepsilon_{\max}$. Thus the inverse solution (5.4) extends to $\varepsilon = \varepsilon_{\max}$. It is now obvious from (5.1) that $\varepsilon_{\max}$ is indeed the maximum of the direct solution $\varepsilon = \varepsilon(\tau)$, for the branch characterized by the given constant $A$. Moreover, for $m > 0$, we may take $\varepsilon_{\text{ref}} = 0$, hence (5.5) becomes

$$F(\varepsilon) \equiv \int_0^{\varepsilon} \frac{d\xi}{\sqrt{A - \alpha \xi^m}}. \qquad (5.9)$$

The function $F$ is one-to-one and increasing in the domain $[0, \varepsilon_{\max}]$, with $F^{-1}(0) = 0$. From (5.4), it is seen that the maximum density was reached at cosmic time

$$\tau_1 = \tau_0 + F(\varepsilon_0) - F(\varepsilon_{\max}) < \tau_0, \qquad (5.10)$$

and *we shall set $\tau_1 = 0$ henceforth*. Due to eq. (5.4) with $\varepsilon_{\text{ref}} = 0$ and $\tau(\varepsilon_{\max}) = 0$, we have

$$\tau = \tau_2 - F(\varepsilon) \quad \text{for} \ \ 0 \leq \tau \leq \tau_2 \equiv F(\varepsilon_{\max}). \tag{5.11}$$

The solution for earlier times $\tau < 0$ is obtained by symmetry with respect to the $\tau = 0$ axis [*cf.* eq. (5.3)], thus

$$\tau = F(\varepsilon) - \tau_2 \quad \text{for} \ \ 0 \geq \tau \geq -\tau_2. \tag{5.12}$$

So $\varepsilon = F^{-1}[\tau_2 \mp \tau]$ cancels as $\tau$ tends towards $\pm \tau_2$. Simultaneously, the scale factor $R$ tends towards $+\infty$ [eq. (4.21)]. Thus, in that cosmology, expansion is accelerated in such a dramatic way that it leads to an infinite dilution in a finite (cosmic, *i.e.* physical) time in the future. Symmetrically, expansion must have been preceded by a contraction starting from an infinite dilution at a finite time in the past. In the case $m = 0$ (*i.e.* $n = 9/2$), eq. (5.9) is replaced by

$$F(\varepsilon) \equiv \int_0^\varepsilon \frac{\mathrm{d}\xi}{\sqrt{\alpha \mathrm{Log}(\varepsilon_{\max}/\xi)}}, \tag{5.9}'$$

which is finite for $0 \leq \varepsilon \leq \varepsilon_{\max}$. The same behaviour is obtained, with eqs. (5.10)-(5.12) still true.

We have for any $n \neq 9/2$, from (4.11) and (4.12): $dt/d\tau = R^n$, hence by (4.21) and (5.3):

$$dt/d\varepsilon = \pm [\varepsilon^{(0)}]^{n/3} \varepsilon^{-n/3} (A - \alpha \varepsilon^m)^{-1/2}. \tag{5.13}$$

If $m > 0$ (*i.e.* $n < 9/2$), this is equivalent to $\pm B\varepsilon^{-n/3}$ as $\varepsilon \to 0$. It follows that, for $n < 3$, both the contraction and the expansion phase take a finite interval of time also if the absolute time is considered − the same interval for both phases. In contrast, if $3 \leq n < 9/2$, one contraction-expansion cycle takes all values of the absolute time $t$ with $-\infty < t < +\infty$, whereas it takes a finite interval of the cosmic time $\tau$. The same is true for $n = 9/2$. This is not satisfying: it means that the theory is unable to predict what happens for values $\tau < -\tau_2$ and $\tau > \tau_2$ for the cosmic, physical time. That situation seems rather unphysical and perhaps should be excluded. (In GR, one is similarly unable to predict what happened before the big bang. But, according to GR, the physical time does not obey the common sense and one is prepared to worse situations, *e.g.* time travels.) Thus, a value of $n$ with $3 \leq n \leq 9/2$ leads to an unpleasant situation.

Let us discuss briefly what happens after the dilution at time $\tau_2 = F(\varepsilon_{\max})$, in the case $n < 3$ (since the theory cannot tell what happens then, if $3 \leq n \leq 9/2$). The mass-energy density $\varepsilon$, relative to the *physical* volume measure $dV$, cancels at $\tau_2$. This is due to the fact that the « absolute » size of the physical length standards (*i.e.,* the size relative to the metric $\mathbf{g}^0$) vanishes as $\tau \to \tau_2$. The mass-energy density relative to the invariable Euclidean volume measure $dV^0$ is, however, constant [eq. (4.23)]. In other words: energy is conserved really. Because $\varepsilon$ cancels at time $\tau_2$ and may not become negative, the rate $\dot{\varepsilon}$ cannot be continuous: it must jump from the value $\dot{\varepsilon}^- = -\sqrt{A}$ to some positive value $\dot{\varepsilon}^+$. (Obviously, $\varepsilon$ must be continuous.) Thus, expansion accelerated to infinite dilution instantaneously gives way to contraction. Due to that discontinuity in $\dot{\varepsilon}$, the solution of (4.27) for $\tau > \tau_2$ represents a new branch, for which $\dot{\varepsilon}^+$ has no relation to $\dot{\varepsilon}^-$, so that the new value $A^*$ of the constant (5.1) is fully *arbitrary*. But the initial data $\varepsilon(\tau_2) = 0$ and $\dot{\varepsilon}(\tau_2) = \sqrt{A^*}$ determines a unique solution of the second-order equation (4.27) in some interval on the right of $\tau_2$. Actually we have explicitly, until the next maximum of $\varepsilon$,

$$\tau - \tau_2 = F^*(\varepsilon) \tag{5.14}$$

where $F^*$ is given by (5.9) with $A^*$ in the place of $A$. After that new maximum $\varepsilon^*_{\max} = (A^*/\alpha)^{1/m}$, the solution is determined by symmetry with respect to the $[\tau = \tau_2 + F^*(\varepsilon^*_{\max})]$ axis, etc.

According to eq. (4.21), the relation $\varepsilon$ - $R$ remains determined by $\varepsilon^{(0)}$, which is a true constant (*i.e.,* valid for all cycles. That $\varepsilon^{(0)}$ must remain unchanged for the next cycle, whereas $A$ may change and thus, *a priori,* will do so, is due to the fact that the constancy equation (4.23) for $\varepsilon^{(0)}$ does not contain the derivative $\dot\varepsilon$, which is discontinuous at the points where $\varepsilon$ cancels). Thus, we have an *aperiodic cyclic Universe with conserved energy.*

### 5.2 Case $n > 9/2$ : one contraction-expansion cycle with non-singular bounce, or expansion from infinite density at past infinity

In this case ($m < 0$), we have $\alpha < 0$ from (5.2), hence $A$ may be positive, negative, or nil [eq. (5.1)]. **If $A < 0$**, then $A/\alpha$ is positive, hence, as before, the radicand in (5.5) does cancel for the value (5.6). Similarly to the case $m > 0$, we have in units such that $A = \alpha = -1$:

$$d\tau/d\varepsilon = \pm [-1 + (1-x)^m]^{-1/2} = \pm [-mx + O(x^2)]^{-1/2}, \qquad x = 1 - \varepsilon \to 0^+. \tag{5.15}$$

Then, *exactly* the same reasoning as for $m > 0$ holds true and eqs. (5.9)-(5.13) are obtained, thus we have a contraction-expansion cycle with finite maximum density, beginning and ending with infinite dilution at finite cosmic times. **If $A \geq 0$**, we have from (5.3):

$$d\tau/d\varepsilon \sim \pm (-\alpha)^{-1/2} \varepsilon^{-m/2} \quad \text{as } \varepsilon \to 0^+, \tag{5.16}$$

and since the r.h.s. is integrable in intervals [0, $\varepsilon$], it follows that here again the infinite dilution shall be reached in a finite cosmic time from now (since we are living in an expansion, we consider the minus sign above), thus we may again take $\varepsilon_{\text{ref}} = 0$ and eq. (5.9) holds true. But now the radicand in (5.9) is defined for whatever value $\xi > 0$, hence arbitrarily large densities have been reached, at past cosmic times defined by eq. (5.4) where $F$ is given by (5.9). From (5.3), we get:

$$d\tau/d\varepsilon \sim -A^{-1/2} \text{ as } \varepsilon \to +\infty \quad (A > 0), \tag{5.17}$$

$$d\tau/d\varepsilon = -(-\alpha)^{-1/2} \varepsilon^{-m/2} \quad (A = 0), \tag{5.18}$$

hence

$$\tau \to -\infty \text{ as } \varepsilon \to +\infty. \tag{5.19}$$

Thus, if $n > 9/2$ and $A \geq 0$, the Universe has been in expansion since an infinite time, and the density increases continuously (with a linear asymptotic behaviour, if $A > 0$), as one looks back in time.

However, eq. (5.13) shows that, if $n > 9/2$, the infinite density has been reached (in the case $A \geq 0$) at a finite *absolute* time in the past, say $t_\infty > -\infty$, even though it is only reached asymptotically as the « physical » (*cosmic*) time $\tau$ tends towards $-\infty$. What could be the physical meaning of events occuring as the absolute time goes beyond the domain corresponding to the whole possible domain of the physical time? But the situation is worse as regards the times at which the infinite dilution ($\varepsilon = 0$) is got, for it is then the exact contrary: from eq. (5.13), it follows that, independently of the value of $A$ and its sign, the infinite dilution is reached only asymptotically as the absolute time $t$ tends towards $+\infty$ (and also as $t \to -\infty$, if $A < 0$), whereas it is reached at a finite value $\tau_2$ of the cosmic time (and also at $-\tau_2$, if $A < 0$). Thus an unpleasant situation occurs for $n > 9/2$, as it also occurs for $3 \leq n \leq 9/2$: the theory is unable to tell what will happen for $\tau > \tau_2$ (and what did happen for $\tau < -\tau_2$, in the case $A < 0$). For this reason, we shall confine the quantitative assessment of time scales to the case $n < 3$.

### 5.3 Comments on accelerated expansion and infinite dilution

At this point one may be tempted to ask what is the physical reason of that seemingly strange and new behaviour predicted for a homogeneous universe by the ether theory, namely an expansion that is accelerated so strongly as to lead to an infinite dilution in a finite time. It should first be remembered that, in the homogeneous universe considered here, matter is subjected to zero external force: the gravity acceleration **g** is identically zero, and no nongravitational force has to be considered (all nongravitational forces are implicitly included in the mass tensor, *e.g.* pressure forces for a perfect fluid). For instance, for a dust (non-interacting matter) in a homogeneous universe, one may check directly that Newton's second law (2.16) with **g** = 0 and $\mathbf{F}_0 = 0$ implies the conservation of momentum in the form (4.19). This is *a priori* obvious anyway since, for a dust, the general equation for continuum dynamics (2.18) is equivalent to (2.16). [36] In the same way, if a dust is considered, the conservation of the material mass-energy (4.21) [or more explicitly (4.23)] is a consequence of Newton's second law (2.16) in a homogeneous universe. Thus there is no need for any force to explain the accelerated expansion undergone by *matter* (which is predicted for whatever behaviour of matter, *i.e.* for any form of the mass tensor), since for a dust it is a direct consequence of Newton's second law with *zero* force. In other words, matter simply follows the accelerated expansion of the preferred frame, *e.g.* it remains at rest in the ether if it was once, and naturally this does not need any force.

So the question is in fact: what is the physical reason that causes the accelerated expansion of the space (preferred frame) itself ? [Remind that this expansion occurs merely in terms of the physical space metric **g** (which is indeed the one to be used in dynamics, see § 2.4), while, in terms of the Euclidean metric $\mathbf{g}^0$, the preferred frame remains perfectly rigid.] The accelerated expansion is surely a physical effect, which is a derivable consequence of eq. (4.27). Now eq. (4.27), and thus this behaviour, is itself a necessary consequence of the theory of gravitation considered here, when applied to a homogeneous universe (although cosmological considerations in the construction of the theory reduce to assuming the flat « background space » (M, $\mathbf{g}^0$) and the absolute time *t*). In particular, the positive sign of the coefficient of $\varepsilon$ in (4.26) [and the coefficient of $\rho_e$ in (4.13)], which is responsible for the truly oscillatory nature of eqs. (4.26) and (4.27) [and already of (4.13)], comes from the equation for the field of ether pressure (2.12), as applied to a spatially uniform field $p_e$. In turn, the form of eq. (2.12) comes from imposing that NG is recovered in the static limit as eq. (2.4), and from the study of small-amplitude, acoustic-like oscillations of the macro-ether relative to its time-averaged position [the latter defining the preferred frame, see eq. (2.2)], that leads to eq. (2.4 bis). [4] Thus, the positive sign and the accelerated expansion come in last resort from the assumption that gravity is like Archimedes' thrust in an elastically compressible fluid. But we must note that, when the accelerated expansion is carried until an infinite dilution is reached, one cannot speak of small oscillations any more. Moreover, at the instant when the zero density is reached, there is strictly speaking no elasticity any more. Hence one might say that, at the points of infinite dilution, the theory is pushed beyond its limits of validity. It seems that this occurs at some place for most theories of gravitation as they are applied to cosmology. For instance, it is obvious that an infinite density makes no physical sense, and also that no equation can apply then: yet a number of theories (including GR) do predict states with infinite density. The present theory avoids infinite densities and, as this curious prediction is got (an infinite dilution in a finite time), it gives the warning that the equations of the theory are used in a situation which is beyond the framework where they were derived. However, since the latter derivation is not a rigorous one (because *a new theory cannot be really derived from an older one*), we have ultimately none other than the equations of the theory, and we should consider their full implications until a square contradiction with observation is found. A nil density seems « less impossible » than an infinite one anyway.

# 6. ESTIMATES FOR SOME TIME SCALES

## 6.1 General case for the contracting-expanding Universe with bounded density ($n < 3$)

Let us evaluate the interval of cosmic time for our contraction-expansion cycle,

$$T = 2 \, F(\varepsilon_{max}), \qquad (6.1)$$

and the interval of cosmic time,

$$\delta\tau_{dilu} = F(\varepsilon_0), \qquad (6.2)$$

that it remains from now (according to this model) before the « infinite dilution » $R = +\infty$. We go over to units in which $A = 1$ and $\alpha = 1$, so that $\varepsilon_{max} = 1$ and $F$ [eq. (5.9)] becomes the dimensionless function

$$f_n(x) \equiv \int_0^x d\xi \, / \sqrt{1 - \xi^{3 - 2n/3}} \, . \qquad (6.3)$$

We obtain for $-\infty < n < 3$, using the MAPLE software,

$$1 < f_n(1) < f_3(1) = 2, \quad \left( f_0(1) = \frac{2}{9} \frac{\pi^{3/2} \sqrt{3}}{\Gamma(2/3)\Gamma(5/6)} \approx 1.40218, \, f_1(1) \approx 1.50055 \right), \qquad (6.4)$$

and the values of $f_n(x)$ for an arbitrary $x$ can be obtained by numerical integration. However, when $x$ is small, we have for $-\infty < n < 3$:

$$f_n(x) \approx x \qquad (x \to 0^+) \qquad (6.5)$$

to a very good approximation: e.g. in the case $n = 1$, we get

$$f_1(x) = x + (3/14) \, x^{10/3} + O(x^{17/3}) \approx x \qquad (x \to 0^+).$$

A straightforward dimensional analysis shows that ($A = 1$ and $\alpha = 1$) is obtained by changing the time unit $[T]$ to $[T]' = \theta \, [T]$ and the density unit $[X] = [M][L]^{-3}$ to $[X]' = \xi \, [X]$, with

$$\xi = \varepsilon_{max}, \qquad \theta = \varepsilon_{max} / \sqrt{A}, \qquad (6.6)$$

where the right-hand sides are evaluated in the starting units. Thus, coming back to those starting units, we find from (6.1) and (6.2):

$$T/2 = \theta f(1), \qquad \delta\tau_{dilu} = \theta f(\varepsilon_0/\varepsilon_{max}). \qquad (6.7)$$

Moreover, extracting $\alpha$ from eq. (5.6) and reinserting in the definition (5.1) of $A$, we get:

$$A = 9\varepsilon_0^2 H_0^2 \left( 1 + \frac{1}{(\varepsilon_{max}/\varepsilon_0)^m - 1} \right), \qquad (m = 3 - 2n/3), \qquad (6.8)$$

whence from (6.6):

$$\theta = \frac{\varepsilon_{max}}{\varepsilon_0} \frac{1}{3H_0} \Big/ \left( 1 + \frac{1}{(\varepsilon_{max}/\varepsilon_0)^m - 1} \right)^{1/2}. \qquad (6.9)$$

With (6.7) and (6.9), it is easy to compute $T$ and $\delta\tau_{dilu}$ as functions of merely the density ratio $\varepsilon_{max}/\varepsilon_0$ and the Hubble constant $H_0$. Let us assume that $\varepsilon_{max}/\varepsilon_0$ is large [and $n < 3$, thus $m > 1$, in order to avoid the unpleasant situation described after eq. (5.13)]. We get to a good approximation:

$$\theta \approx \frac{\varepsilon_{max}}{\varepsilon_0} \frac{1}{3H_0} \equiv \theta_{approx} \qquad (n < 3 \text{ and } \varepsilon_{max}/\varepsilon_0 \gg 1). \qquad (6.10)$$

It follows by (6.5) and (6.7) that, for $n < 3$ and $\varepsilon_{max}/\varepsilon_0 \gg 1$,

$$\delta\tau_{dilu} \approx 1/(3H_0) \approx 3.27 \, h^{-1} \times 10^9 \text{ years} \qquad (6.11)$$

where

$$H_0 = 100 \, h \text{ km/s/Mpc}, \qquad 0.50 \leq h \leq 1$$

(recall that 1 Mpc $\approx 3.09 \times 10^{19}$ km). We get further for $n < 3$ and $\varepsilon_{max}/\varepsilon_0 \gg 1$:

$$T/2 \approx (\varepsilon_{max}/\varepsilon_0) f_n(1) \, \delta\tau_{dilu} \approx 3.27 \, h^{-1} f_n(1) \, (\varepsilon_{max}/\varepsilon_0) \times 10^9 \text{ years}, \qquad (6.12)$$

with $1 < f_n(1) < 2$ by (6.4), and even $1.4 < f_n(1)$ if $n \geq 0$. Thus, in the « pleasant » domain $n < 3$, the estimates do not depend much on $n$. Note that $T/2$ is the duration of the expansion phase, *i.e.*, the time interval from the maximum density to the infinite dilution, so that the time elapsed from the maximum density to now (the « age of the Universe ») is:

$$T' = T/2 - \delta\tau_{dilu}.$$

If one assumes that the 2.7 K radiation is a relic of a high density state, and if one neglects the interaction of that radiation with matter, then the scale factor must have been reduced by a factor of approximately 1500 since the « decoupling, »[1, 3] thus

$$(\varepsilon_{max}/\varepsilon_0) \geq 1500^3 = 3.375 \times 10^9 \Rightarrow \quad T/2 \geq 1.1 \, h^{-1} \times 10^{19} \text{ years} \qquad (n < 3). \qquad (6.13)$$

Nearly that huge number is then got for the « age of the Universe » $T'$, too. In that case, galaxies have much time to form, but may be such a huge time poses different problems. We mention that very similar estimates are obtained in the unpleasant case $3 \leq n \leq 9/2$.

**6.2. Redshifts and the case $n = 0$ (no cosmological time-dilation)**

The special feature of the case $n = 0$ is eq. (5.7), that expresses the density ratio $\varepsilon_{max}/\varepsilon_0$ as function of the current energy density $\varepsilon_0$ and the Hubble parameter $H_0$. Since

$$G \approx 6.67 \times 10^{-8} \text{ cm}^3 \text{ g}^{-1} \text{ s}^{-2},$$

we get with $H_0 \approx 50$ km/s/Mpc and $\varepsilon_0 \approx 10^{-30}$ g/cm$^3$ :
$$\varepsilon_{max}/\varepsilon_0 \approx 3.51. \qquad (6.14)$$

Changing the values of $H_0$ and $\varepsilon_0$ within observational limits does not change a lot this ratio, *e.g.*

$$\varepsilon_{max}/\varepsilon_0 \approx 21.2 \qquad (6.15)$$

with $H_0 = 75$ km/s/Mpc and $\varepsilon_0 = 10^{-32}$ g/cm$^3$. Thus, if $n = 0$, one has a cosmological model predicting a maximum density only a few times larger than the current mean density in the Universe. *I.e.,* a state with high density has not been reached in that case. This means that the big bang cosmology and hence the currently accepted explanation of the cosmic microwave background and the abundances of the light elements do not hold true if $n = 0$. This kind of « non-singular bounce » with low density ratio is predicted by Rastall's theory[54] and in one case for Rosen's theory. [55] In our opinion, this prediction *per se* does not invalidate those cosmologies, even less the underlying theories of gravitation. In order to explain the production of the light elements and the microwave background, an explanation of the kind advocated by Hoyle *et al.*,[56] [involving little big bangs distributed over space and time], might also be appropriate (see Section 7). [10] However, one serious difficulty for models that predict a low density ratio is the explanation of high redshifts $z$. [57] Indeed, one has generally, for any comoving object

$$z = R(\tau_0)/R(\tau_{emission}) - 1 = [\varepsilon(\tau_{emission})/\varepsilon(\tau_0)]^{1/3} - 1 \leq (\varepsilon_{max}/\varepsilon_0)^{1/3} - 1. \qquad (6.16)$$

Thus, in addition to the fact that the assumption of a cosmological time-dilation ($n \neq 0$) fits better with the rest of the present ether theory, it seems that the case $n = 0$ is ruled out by the observation of galaxies exhibiting a cosmological redshift $z \approx 4$, hence $\varepsilon/\varepsilon_0 \approx 100$. From (5.11) and (6.6), it follows that, if $n \leq 9/2$, the time elapsed since light was emitted from a comoving galaxy, as the density was $\varepsilon$, is

$$\delta\tau = \theta \, [f(\varepsilon/\varepsilon_{max}) - f(\varepsilon_0/\varepsilon_{max})] = \theta \, \{f \, [(\varepsilon/\varepsilon_0) \times (\varepsilon_0/\varepsilon_{max})] - f(\varepsilon_0/\varepsilon_{max})\}. \qquad (6.17)$$

Hence, by (6.9) and the equality (6.16), we may calculate this time once we know $H_0$, the redshift $z$, and the density ratio $\varepsilon_{max}/\varepsilon_0$. If $\varepsilon/\varepsilon_{max}$ is small and if $n < 3$, as it should be in order to avoid unpleasant situations [see after eq. (5.13)], we may evaluate this easily as

$$\delta\tau \approx [(1+z)^3 - 1]/(3 \, H_0), \qquad (6.18)$$

which also provides, even if $\varepsilon/\varepsilon_{max}$ is not small, an order-of-magnitude estimate. Thus, light emitted by a comoving galaxy and received at $z = 1$ has been emitted, according to that theory (in the case of the bouncing universe, *i.e.* $n < 3$), some $23 \, h^{-1} \approx 30$ billions of years ago.

## 7. DISCUSSION

It has been examined what are the main features of the cosmology predicted by a very simple theory of gravitation, which is explicitly an ether theory and thus gives a physical role to the *vacuum*. Such role is now to be expected for several reasons, one of which coming from observational data relevant to cosmology. Indeed, expansion is now found to be accelerated and thus, within general relativity (GR), it becomes necessary [see eq. (4.16)] to introduce a positive cosmological constant in order to fit the observations – [9-10] thereby giving a physical role to the *vacuum*. The most important result is that the ether theory predicts that expansion *must* be accelerated, indeed strongly [eq. (4.15)]. It may seem that this predicted acceleration is too strong since the current data seem best fitted with $q \approx -1$ rather than with $q \approx -4$. [9-10] However, it should be realized that (i) according to Riess *et al.*[9] the current data, although they already seem to exclude a positive value of $q$, are not yet enough to conclude as to the value of $q$ (in particular, it is desirable that luminosity distances of supernovae at higher redshifts may be included in future work); and (ii) the fitting is done *within GR* and depends further on the particular general-relativistic cosmological model. For instance, when doing the fitting, Riess *et al.*[9] exclude models giving rise to a "bounce" without a big bang, whereas such non-singular bounce is quite likely according to the ether theory. In a future work, it would be worth to investigate the luminosity distance - redshift relation within the present model and thus to check whether this model can quantitatively account for redshift observations. It is also worth to recall that $q = -1$ is simply the value predicted for a Universe expanding *at a constant rate* ($H$ = Const): thus, the current evidence for an accelerated expansion would seem *a priori* (*i.e.,* independently of any model) favour $q < -1$. [11] In our opinion, the most important point is that the assumption of a homogeneous universe is very strong. Even if the universe is really homogeneous on a large scale (which seems plausible), one should expect that its observed inhomogeneity does influence the effective values of the parameters entering a homogeneous model, such as $H$ and $q$. These have indeed the status of « effective macroscopic parameters » in the sense of the homogenization theory. Thus one should expect that those values of the effective parameters that are obtained under the simplest model (which neglects the inhomogeneity) are not the correct ones. In view of this discussion, it seems important that the present model predicts unambiguously what now appears to be the correct tendency, *i.e.* accelerated expansion.

Depending on whether the cosmological time-dilation exponent $n$ [eq. (4.11)] is larger or smaller than 9/2, and depending on an integration constant $A$ [eq. (5.1)], the present theory predicts *two kinds of evolutions for the Universe:*

(**i**) The first kind ($n > 9/2$ and $A \geq 0$) is an expanding universe with unbounded density: as the cosmic time increases infinitely in the past, the density increases without limit. This case provides thus arbitrarily high density in the past, without a big-bang singularity. However, it has two curious features (see the end of § 5.2): *a)* the absolute time of the theory tends towards a finite value $t_\infty$ while the physical, cosmic time $\tau$ tends towards $-\infty$ ; *b)* the infinite dilution shall be reached at a finite cosmic time $\tau_2$, but this will take an infinite amount of the absolute time $t$ entering the equations of the theory, so that nothing can be predicted for $\tau > \tau_2$. This latter feature is true for any $n \geq 3$, independently of $A$, and seems rather unphysical, hence the case $n \geq 3$ has not been presented here at the stage of time-scales evaluation.

(**ii**) The second kind [$n \leq 9/2$, or ($n > 9/2$ and $A < 0$)] is a set of symmetric contraction-expansion cycles: there is one such cycle, if $3 \leq n \leq 9/2$, as also if ($n > 9/2$ and $A < 0$), and there is an infinite sequence of them, if $n < 3$. Each cycle starts from an infinite dilution: space contracts until the energy density reaches a finite maximum, after which expansion takes place until an infinite dilution is obtained. Thus each cycle contains a « non-singular bounce, » as is also predicted by several other bimetric theories of gravitation when the background metric is flat[54, 55, 58]. In Rastall's theory, [54] and in one case for Rosen's theory, [55] the maximum density is only a few times higher than the current density. If one would assume no cosmological time-dilation ($n = 0$), this would be also the case for the present theory (§6.2). This seems to be ruled out by the observation of high cosmological redshifts. In addition, assuming a cosmological time-dilation is more consistent with the spirit of the present ether theory. When one indeed assumes this, the ratio of the maximum density to the current density, $\varepsilon_{max}/\varepsilon_0$, is not constrained by the cosmological model. This also happens in Petry's theory. [58] The « age of the Universe » is then determined by the data of the Hubble parameter $H_0$ and the ratio $\varepsilon_{max}/\varepsilon_0$. If one admits the standard explanation of the 2.7 K radiation as resulting from a previous stage with a very high density, one then gets a huge value for the « age of the Universe » [eq. (6.13)]. Even for a maximum density $\varepsilon_{max}/\varepsilon_0 \approx 100$ [which seems to be the minimum required to interpret the *observed* redshifts as cosmological, eq. (6.16)], the age is very large, several hundreds of billions of years [eq. (6.12)]. Both results are true independently of the value $n$ with $n < 3$ (and even $n \leq 9/2$; no estimate has been done for $n > 9/2$).

Thus, the present theory proposes two possible scenarios in which high density has been reached in the past: either an expanding universe with infinitely increasing density as the cosmic time goes back to infinity [although this case has an unpleasant feature, see point (**i**) *b*) above], or bouncing universes with high density ratio. Now, despite the success of the big-bang models, we are currently not *certain* that the Universe did really pass through a state of very high density in the past. The big-bang models predict a singularity with infinite density, which can hardly be considered plausible. Further, the big-bang models are based on GR and so can cope with the currently plausible acceleration of the expansion only by assuming a positive cosmological constant [see eq. (4.16)]. But this produces a large-scale repulsive force that seems to pose serious problems at the scale of galaxy clusters.[59] May be one day it shall be proved that a very high density has been reached in the past, *e.g.* from the observation of very high redshifts proved to be cosmological. Currently, it seems that alternative proposals for the origin of the microwave background and the light elements should be considered also. In the quasi-steady-state model, proposed by Hoyle *et al.*,[56] matter creation would occur in the form of bursts affecting very massive objects (roughly $10^{16}$ solar masses) with high gravitational fields. Thus, in this model, matter creation occurs in localized regions of space-time, but the distribution of those regions should be macro-uniform, whence the « quasi-steady » character of that cosmology. According to the investigated ether theory, matter creation or destruction *must* occur if we have a fluid pressure and a variable gravitational field.[6, 36] Specifically, *creation* indeed occurs in an expanding fluid ball with a high density.[36] Moreover, the gravitational collapse does lead to such expanding balls (after a non-singular bounce, again).[27] Thus, the quasi-steady-state model is compatible with the ether theory considered in the present work. As stated by Hoyle *et al.*,[56] the consideration of such expanding « fire balls » with matter creation could provide, not merely an alternative explanation for the microwave background and the light elements production (as described in detail in Ref.[56]), but also an understanding of the physics of active galactic nuclei and gamma-ray bursts (see also Brandes[60] as regards the latter point). Much work would still be needed to confirm or invalidate this idea, even though important work has already been done in that direction by Hoyle and his coworkers.[56] At the present stage of investigation of the cosmological models in the ether theory, it is not possible to state which cosmological scenario should apply to our Universe, *i.e.*, whether it is a mere expansion with the density infinitely increasing at past infinity ($n > 9/2$ and $A \geq 0$), or one contraction-expansion cycle with bounce [$3 \leq n \leq 9/2$, or ($n > 9/2$ and $A < 0$)], or an infinite sequence of such cycles ($n < 3$). One just may say that the latter case is preferable, because an unpleasant situation occurs for $n \geq 3$. In the case $n < 3$, the density ratio of the current cycle seems to be constrained only to be greater than 100, which is far too small to decide between scenarios invoking a global high density state in the past, and the quasi-steady-state scenario.

Among cosmologies predicting a contraction-expansion cycle with non-singular bounce, the original feature of the present cosmology (for $n < 3$) is that both the contraction and the expansion phase take a *finite time* (the same time for both phases): it predicts that in the future, indeed in some 6 billions of years from now, the expansion of the Universe will accelerate to the point of giving an infinite scale factor. (Recall: this is obtained under the usual oversimplifying assumption of a homogeneous universe. Anyway, this will not be the end, since energy is conserved and a contraction phase will immediately follow.) There are other theories which predict a cyclic universe, however, *e.g.* GR with a negative value of $\Lambda$.[2] In Kaya's model,[61] a succession of identical *expansion-contraction* cycles is predicted, each cycle beginning and ending with infinite density (big-bang and big-crunch), thus the same scenario as in GR with $\Lambda < 0$, and a quite different one from those predicted here for $n < 3$. Kaya's model modifies Newton's theory of gravitation only by the introduction of a time-varying gravitation « constant » $G$, but it also gives physical properties to *vacuum*. Namely, Kaya's « energy-matter space » is assumed to be a ball endowed with a kind of radial elasticity, and it stores elastic energy as it expands isotropically. This explains why expansion is gradually braked and after some stage gives way to contraction in his analogical model, and this in a periodic way (« like a giant pendulum »). In the present model, we have seen in § 5.3 that the (nonlinear) *oscillatory* behaviour of the Universe is inherited from eqs. (2.3), (2.4) and (2.4 bis) of the construction stage of the theory, that express gravity as Archimedes' thrust in the elastically compressible micro-ether. Hence the oscillatory behaviour of the Universe may be considered to follow, in both models, from the elasticity of *vacuum:* in the case of Kaya's analogical model of the universe as a ball endowed with radial elasticity, this would be a stretched medium, and unable to resist compression – whence the singularity with infinite density. In the model deduced here from the investigated theory of gravitation, *vacuum* would be like a compressible fluid, unable to resist tension: this picture might help to get some representation of the accelerated expansion, currently observed. Kaya's cosmology, and most models based on GR, have singular points, because those models predict states with infinite density. The cosmological model based on the scalar ether theory also has singular points, because it predicts states with infinite dilution, at which points the time-derivative of the mass-energy density is discontinuous. The singular points are thus of a less severe kind for the present model than for models based on GR.

**Acknowledgement.** I am very grateful to Professors Eugen Soós (Bucharest) and Pierre Guélin (Grenoble) for their encouragements to my work on gravitation. In particular, Eugen Soós early suggested that I should investigate cosmology in my theory. I would also like to thank the Reviewers of the present paper for their numerous constructive remarks, that led me to improve the presentation of the theory and the discussion of its cosmological implications.

**Endnotes**

1. « Die Einführung eines "Lichtäthers" wird sich insofern als überflüssig erweisen, als nach der zu entwickelnden Auffassung weder ein mit besonderen Eigenschaften ausgestatteter "absolut ruhender Raum" eingeführt, noch einem Punkte des leeren Raumes, in welchem elektromagnetische Prozesse stattfinden, ein Geschwindigkeitsvektor zugeordnet wird. »

2. This description applies to what would be found by observers near the surface of the collapsing body, using their meters affected by the gravitational field *i.e.* (see §§ 2.3 and 2.4) by the field of « ether pressure » $p_e$. This « bounce in terms of physical meters » is due to the time evolution of $p_e$ in the zone occupied by the body. For the collapse in free fall with spherical symmetry, this evolution is governed by an equation like (4.27) below [although, in the present work, eq. (4.27) is derived for a homogeneous universe]. If observers could use the « Euclidean meters, » not affected by the field $p_e$, they would find instead that the body remains undeformed, *i.e.*, it undergoes neither an implosion nor an explosion. Just the same is also true in the case of a homogeneous universe: its measurable expansion or contraction, predicted by the cosmological model derived from the present theory (§§ 5.1 and 5.2), is detected with physical meters affected by the field $p_e$, and does not occur in terms of the Euclidean « background » space metric. See § 5.3 for comments on the physical cause of the expansion.

3. Latin indices vary from 1 to 3 (spatial indices), Greek indices from 0 to 3.

4. The *spatial* volume average (with respect to the Euclidean volume measure $V^0$), that conceptually determines the « macroscopic » fields $\mathbf{u}_e$, $p_e$ and $\rho_e$, should be done at a large, but finite scale, because the resulting fields still depend on the spatial position (and on the time). The velocity field that conceptually defines the preferred reference frame E, still involves an additional *time*-average, eq. (2.2), whose result is **0**, by definition of E. Actually, the field $p_e$ [or equivalently the associated field $\rho_e = p_e/c^2$, eq. (2.10)] is the gravitational field and is hence operationally determined by solving the independent equations of the proposed theory [eqs. (2.12) and (2.18)], accounting for the initial conditions and the equation of state of the matter tensor **T**.

5. The case where $\nabla \rho_e (t, \mathbf{x}) = 0$ actually occurs at equilibrium points ($\mathbf{g} = 0$), which are isolated. (Here $\nabla$ means indifferently the gradient operator with respect to metric **g** or metric $\mathbf{g}^0$.) Then an isopotential coordinate system may be defined as before, so the form (4.1) for the metric is true except at such isolated points. Hence, eq. (4.2) applies (meaning that measure $V$ on M has density $R^3 \rho_e^\infty / \rho_e$ with respect to measure $V^0$). For the special case that is obtained in homogeneous cosmological models, it is shown in § 4.2 that eqs. (4.1) and (4.2) still apply.

6. At the macroscopic scale. The preferred frame E or is rigid with respect to $\mathbf{g}^0$, and is thought of as following the average motion of a « micro-ether », which would be a perfect fluid (see § 2.2).

7. It is actually sufficient that the homogeneity be reached only asymptotically as the size of the sample of universe tends towards infinity, independently of the spatial position of the centre of the sample. Thus, cosmology would describe only asymptotic averages (stationary in space) of the matter and gravitational fields. The cosmological principle does *not* impose, therefore, that the size of astronomical structures should have an upper bound.

8. In some « Newtonian » cosmologies, a similar formula as eq. (4.16) is obtained [Ref. [(2)], p. 226]. But, in such models, Einstein's concept of local inertial frames is substituted for Newton's concept of global ones [(2)] (in true Newtonian gravity (NG), the matter distribution must be spatially bounded). Yet Newton's third law is essential to NG and asks for global inertial frames. [(4)] In a homogeneous universe, the attraction field should be $\mathbf{g} = 0$, really.

9. It is sufficient to verify this for a dust. We have then $T_0^{\ 0} = \rho^* \gamma_v^{\ 2}$ with $\gamma_v$ the Lorentz factor. Since the proper volume of dust $dV_{\text{pr}}$ is contracted to $dV = dV_{\text{pr}} / \gamma_v$ when it is evaluated with the physical metric

    in the frame E, the rest-mass density with respect to d$V$ is $\rho = \rho^* \gamma_v$, hence the density of rest-mass plus kinetic energy is indeed $\rho \gamma_v = \rho^* \gamma_v^2$.

[10] One also might think that the light elements could have been produced during a foregoing expansion-contraction cycle, this one leading to a high maximum density.

[11] It still may be said that the present work is the simplest cosmological model based on the ether theory, in that it is based on a flat background metric: the ether theory may also be formulated if the invariable space metric $\mathbf{g}^0$ is one with constant curvature. The examination of cosmological models with such non-flat reference metric is left to a future work. However, a preliminary study seems to indicate that they give little or no difference as to the value of $q$.

**Mayeul Arminjon**
Laboratoire « Sols, Solides, Structures »
Institut de Mécanique de Grenoble
B.P. 53, 38041 Grenoble cedex 9, France
E-mail: arminjon@hmg.inpg.fr